\documentclass[fleqn,12pt,twoside]{article}
\usepackage[headings]{espcrc1}

\readRCS
$Id: espcrc1.tex,v 1.2 2004/02/24 11:22:11 spepping Exp $
\ifx\pdftexversion\undefined
  \usepackage[dvips]{graphicx}
\else
  \usepackage[pdftex]{graphicx}
\fi
\newcommand{\AmS}{{\protect\the\textfont2
  A\kern-.1667em\lower.5ex\hbox{M}\kern-.125emS}}

\title{The role of elliptic flow correlations in the discovery of the sQGP at RHIC}

\author{Roy A. Lacey\address[SBChem]{Dept. of Chemistry, 
        Stony Brook University,   
        Stony Brook, NY, 11794-3400, USA.}
        }
       
\begin{document}

\maketitle

\begin{abstract}
	Flow measurements are reviewed with particular emphasis on the   
hydrodynamic character of elliptic flow at RHIC. Hydrodynamic scaling 
compatible with the production of highly thermalized matter having 
a high degree of collective interactions and extremely low viscosity,
is observed for a broad selection of the data. These properties suggest  
the production of a new state of strongly interacting nuclear matter 
at extremely high density and temperature where the relevant degrees 
of freedom are the valence quarks, ie. the sQGP. 
%
\end{abstract}

\section{Prologue}

For more than twenty five years the heavy ion community has sought to create and study 
the quark gluon plasma (QGP) -- a new phase of hot and dense nuclear matter where quarks 
and gluons are no longer confined to the interior of single hadrons. Such a state is 
predicted to exist at high energy densities by quantum chromodynamics (QCD) and is now 
very strongly indicated by lattice QCD calculations~\cite{fodor-katz02,bielefeld-QCD}.
The fundamental value of this search is rooted in the fact that the QGP; (i) is the ultimate 
primordial form of QCD matter at high temperature or baryon number density, (ii) was present during 
the first few microseconds of the Big Bang, (iii) provides an example of phase transitions 
which may occur at a variety of higher temperature scales in the early universe, and (iv) can 
provide important insights on the origin of mass for matter, and how quarks are confined 
into hadrons \cite{Gyulassy_NPA750}. 

\begin{figure}[htb]
\begin{minipage}[t]{0.5\linewidth}
\includegraphics[width=1.\linewidth]{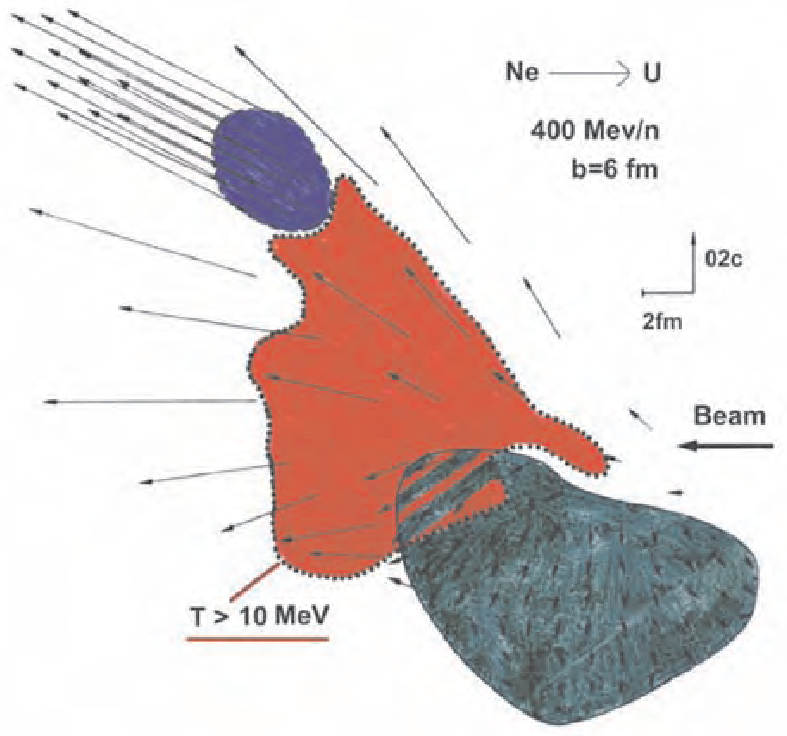}
\vskip -1.2cm
\caption{ Density and temperature contours in the scattering plane of  
a mid-central Ne+U collision. The arrows indicate velocity fields. 
Results are from a fluid-dynamical calculation \cite{Stocker_PRL44}.}
\label{fig:splash}
\end{minipage}
\hskip 0.2cm 
\begin{minipage}[t]{0.5\linewidth}
\includegraphics[width=1.\linewidth]{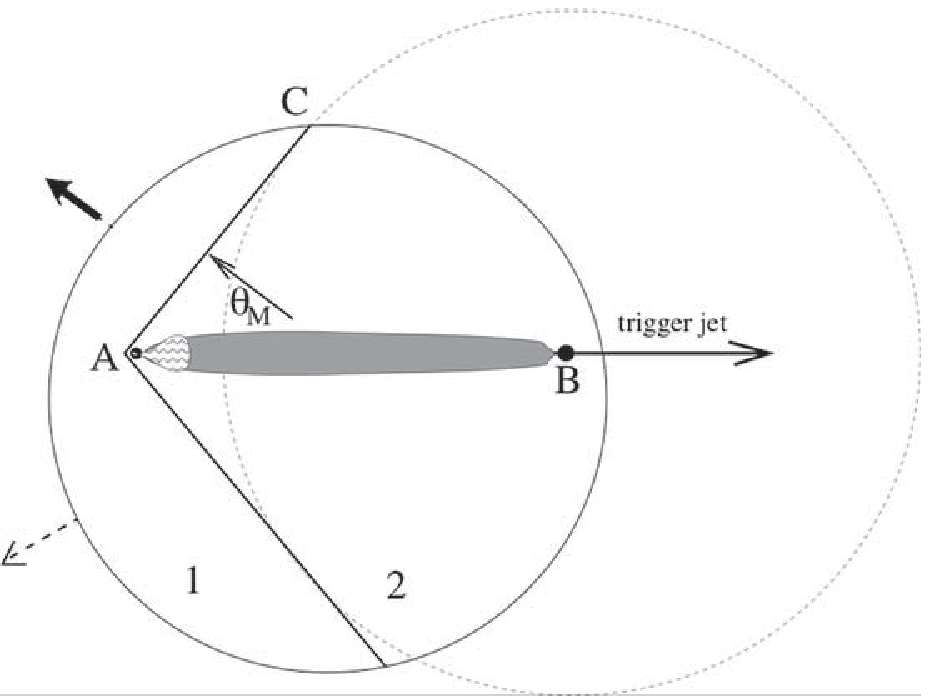}
\vskip -1.2cm
\caption{Schematic diagram of the development of a Mach cone resulting from 
the strong interaction between a jet and a high energy density medium \cite{mach_cones}.
The arrows which point to the left indicate the direction of flow normal to the shock cone. }
\label{fig:mach_cone}
\end{minipage}
\end{figure}

	The use of flow correlations as a probe for the nuclear equation of state (EOS) and 
a possible phase transition, was recognized quite 
early \cite{GlassGold_AP6,Chapline_PRD8,Sheid_PRL32,Stocker_PRL44}. 
The connection is made transparent in the framework of perfect fluid
dynamics where the conceptual link between the conservation laws (baryon number,
and energy and momentum currents) and the fundamental properties of a fluid (its equation of
state and transport coefficients) is straightforward. It is therefore not surprising that 
hydrodynamic considerations have played, and continue to play an important role in flow 
studies \cite{Flow_Reviews}. Fig.~\ref{fig:splash} shows the result of an early fluid-dynamical calculation which 
illustrates the development of flow correlations in Ne+U collisions. Fig.~\ref{fig:mach_cone} 
shows a schematic diagram of a more recent conjecture that jet interaction with a high energy 
density medium could lead to the generation of a Mach cone and hence conical flow \cite{mach_cones}.

\subsection{Azimuthal distributions and flow correlations }
Azimuthal angle distributions and correlation functions play a pivotal role in the study 
of flow correlations.
Experimentally, one measures the magnitude of flow correlations by evaluating 
the Fourier coefficients $\lambda_n$, of the anisotropy of the distribution in 
azimuthal angle difference ($\Delta\phi = \phi_1 - \phi_2$) between pairs of charged 
hadrons \cite{wang91,lacey93,Lacey_QM02}
\begin{equation}
\frac{dN}{d\Delta\phi} \propto 
\sum\limits_{n =  - \infty }^\infty  {\lambda _n } e^{in\Delta \phi }, 
\ \hskip 0.5cm \lambda_n = \langle e^{in\Delta\phi}\rangle.
\label{delta_phi}
\end{equation}
Here $\phi_{1,2}$ are the azimuthal angles of a pair of particles
measured in the laboratory coordinate system, and the brackets denote an 
average over pairs of particles emitted in an event followed by further 
averaging over events. Alternatively, the Fourier coefficients $v_n$ can be obtained 
as \cite{Danielewicz:1985hn,Poskanzer:1998yz,Ollitrault:1993ba,Ollitrault:1997di}:
\begin{equation}
v_n\equiv\langle e^{in(\phi-\Phi_R)}\rangle
=\langle \cos n(\phi-\Phi_R)\rangle,
\label{rxn_plane}
\end{equation}
where $\phi$ is the azimuthal angle of an emitted particle
(also measured in the laboratory coordinate system) and $\Phi_R$ is 
an estimate of the azimuth of the reaction plane. A requisite correction
which takes account of the dispersion of the reaction plane is easy to 
evaluate \cite{Danielewicz:1985hn,Poskanzer:1998yz,Ollitrault:1993ba,Ollitrault:1997di}.  

For detectors having a limited $\phi$ acceptance, 
the correlation function $C_2(\Delta\phi)$, is often exploited to 
measure $\lambda_n$ \cite{wang91,lacey93,Lacey_QM02}:
\begin{equation}
C_2 \left( {\Delta \phi } \right) = \frac{{N_{cor} \left( {\Delta \phi } \right)}}
{{N_{mix} \left( {\Delta \phi } \right)}} = \sum\limits_{ - \infty }^{ + \infty } {\lambda_n e^{in\Delta \phi } },
\label{c2_cor}
\end{equation}
where the correlated distribution $N_{cor}\left( {\Delta \phi }\right)$, is obtained from pair members 
belonging to the same event. The mixed distribution $N_{mix}\left( {\Delta \phi }\right)$,  
is made of pair members belonging to different events having characteristics (multiplicity, vertex, etc.) 
similar to those for $N_{cor}\left( {\Delta \phi }\right)$. 
The correlation between each particle and the reaction plane induces correlations among all particles.
Thus, it is straightforward to show that $\lambda_n = v_n^2$ when both particles are selected 
from the same $p_T$ and rapidity $y$, range (ie. fixed-$p_T$ correlations) and $\lambda_n = v_{n}(1)v_{n}(2)$ 
when particle (1) and (2) are selected from a different $p_T$ and/or y range 
(ie. assorted correlations). When $v_n$ is measured for specific particle species 
or as a function of transverse momentum, centrality 
and/or rapidity, they are referred to as ``differential'' flow measurements. 
When the measurements are averaged over a sizable phase space they are 
termed ``integral" measurements. The first two coefficients, $v_1$ and $v_2$, 
are usually referred to as directed flow and elliptic flow, respectively.

	In addition to the consideration of detector acceptance, reliable flow measurements 
often require the suppression of so called non-flow correlations. These include 
small-angle correlations due to final state interactions and quantum statistical 
effects \cite{Kotte_2005}, correlations due to resonance decays \cite{Hong_phys_lett_b407}
and mini-jet production \cite{Kovchegov_NPA708}. The methods of 
Lee–Yang zeros \cite{Bhalerao_NPA727} and cumulants \cite{Borghini_PRC64} have been developed 
to suppress such non-flow correlations. The cumulant method is based on a cumulant expansion 
of multi-particle correlations \cite{Borghini_PRC64}.
The method of Lee–Yang zeros follows the spirit of the Lee–Yang theory of phase 
transitions \cite{Yang_PR87}, and  extracts flow directly from the genuine 
correlations involving a large number of particles \cite{Bhalerao_NPA727}.

Azimuthal distributions are not only important for flow measurements. They are of great 
current interest in connection with the study of jet modification \cite{Holzmann:2005vq}, 
the formation of disoriented chiral condensates \cite{Nandi:1999vb,Asakawa:2000dr}, and the study of 
parity and/or time-reversal violation\cite{Voloshin:2000xf}. The combined use of flow 
correlations and two-particle interferometry measurements is also used extensively to 
gain detailed insight on the three-dimensional structure of 
emitting sources \cite{Voloshin:1996mc,Wiedemann:1998cr,Heiselberg:1999ik}.

\subsection{A snapshot of over twenty years of elliptic flow measurements }
\begin{figure}[htb]
\begin{minipage}[t]{0.5\linewidth}
\includegraphics[width=1.\linewidth]{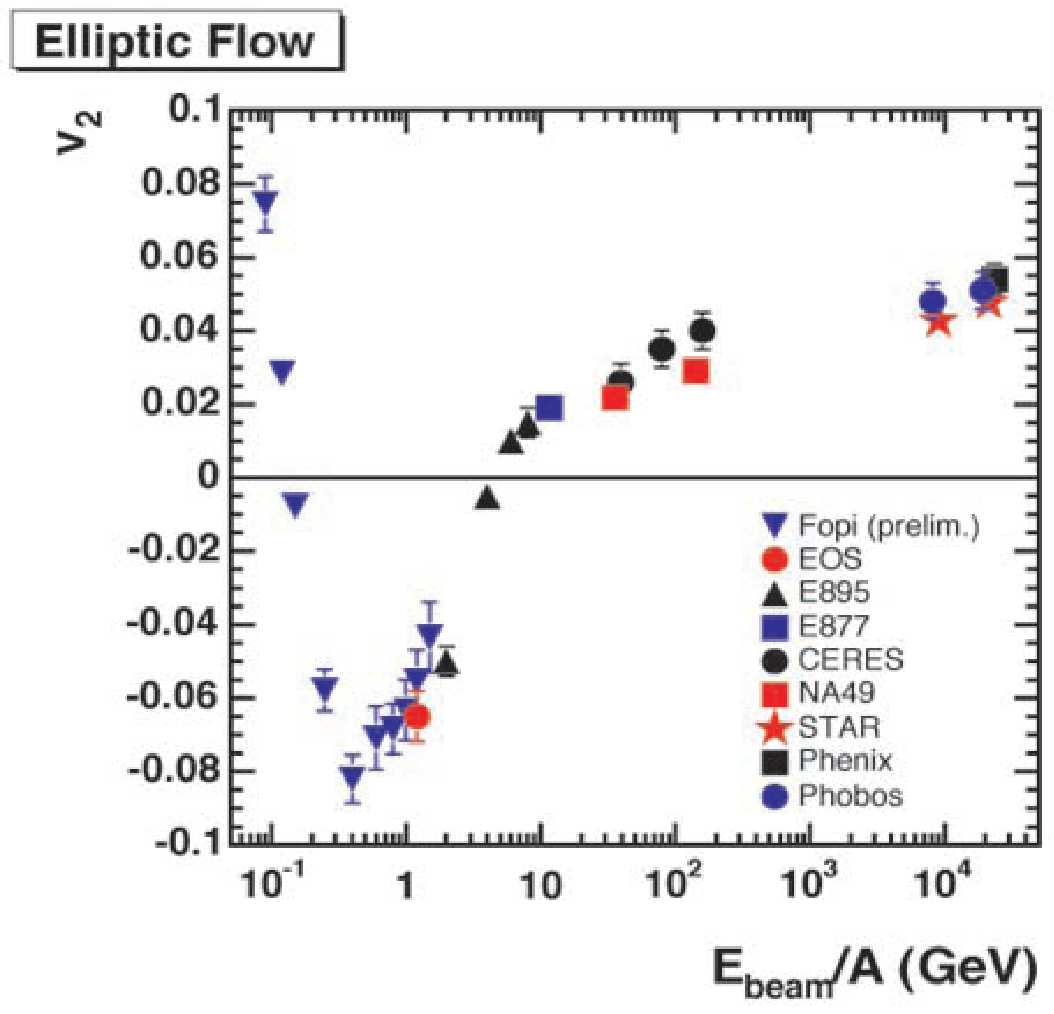}
\vskip -1.2cm
\caption{ Elliptic flow ($v_2$) vs beam energy for mid-central collisions
Au+Au colllisions \cite{Wetzler}.}
\label{fig:elip_flow_systematics}
\end{minipage}
\hskip 0.2cm 
\begin{minipage}[t]{0.5\linewidth}
\includegraphics[width=1.\linewidth]{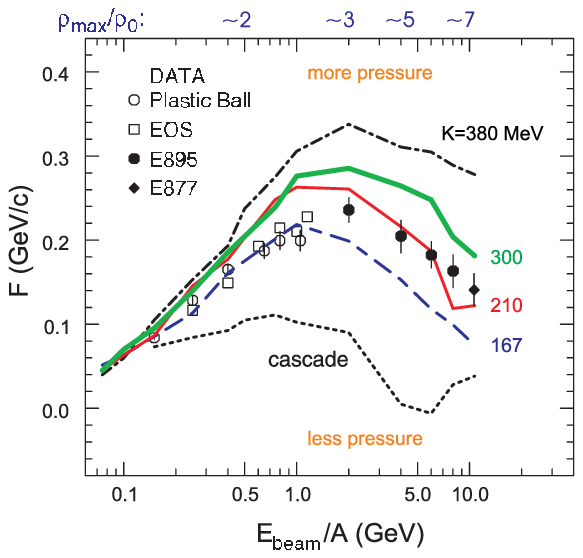}
\vskip -1.2cm
\caption{ Directed flow vs. beam energy for mid-central Au+Au collisions. 
The lines show transport theory predictions for different EOS's \cite{Danielewicz:2002pu}.}
\label{fig:side_flow_systematics} 
\end{minipage}
\end{figure}

Systematic flow measurements now exist for a beam energy range which spans 
nearly six orders of magnitude. They include measurements done at  
MSU, GANIL, BEVALAC, GSI, Dubna, AGS, SPS and RHIC. Results for elliptic  
and directed flow in Au+Au collisions, from several of these measurements are summarized in 
Figs.~\ref{fig:elip_flow_systematics} and \ref{fig:side_flow_systematics} respectively. 
	For  a broad range of beam energies ($E_{beam}/A\geq$~400 MeV), the elliptic flow 
results can be understood in terms of a delicate balance between (i) the ability of pressure developed 
early in the reaction zone, to effect a rapid transverse expansion of nuclear matter, 
and (ii) the passage time for removal of the shadowing of participant hadrons by the 
projectile and target spectators\cite{sor97,dan98,Danielewicz:2002pu}. 
The characteristic time for the development of expansion perpendicular
to the reaction plane is $\sim R/c_s$, where the speed of sound 
$c_s = \sqrt {\partial P/\partial \varepsilon}$, $R$ is the nuclear radius, $P$ is the pressure 
and $\varepsilon$ is the energy density. The passage time is $\sim 2R/\left( {\gamma _0 v_0 } \right)$,
where $v_0$ is the c.m. spectator velocity. 

\begin{figure}[thb]
\begin{minipage}[t]{0.5\linewidth}
\includegraphics[width=1.\linewidth]{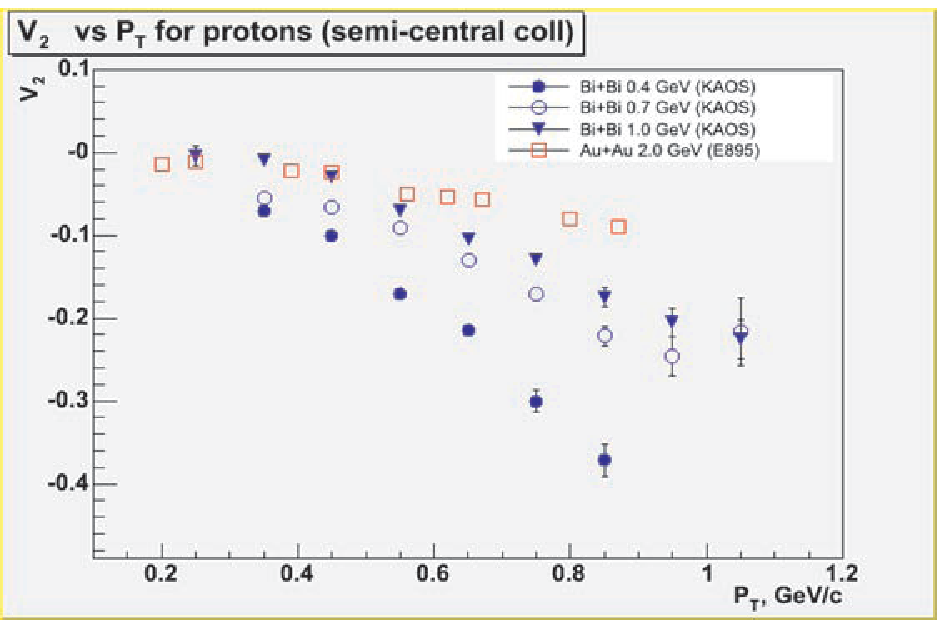}
\vskip -1.0cm
\caption{ Differential flow $v_2(p_T)$ vs $p_T$ for several beam energies.
The data are from Refs.~\cite{kaos,pin99}  
}
\label{fig:kaos_unscaled_v2}
\end{minipage}
\hskip 0.2cm 
\begin{minipage}[t]{0.5\linewidth}
\includegraphics[width=1.\linewidth]{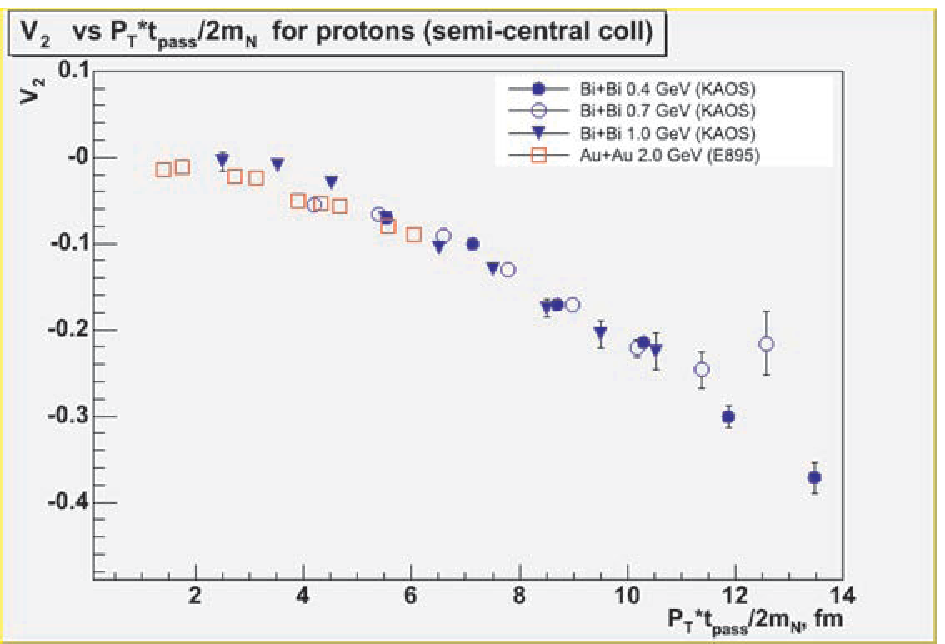}
\vskip -1.0cm
\caption{ The same data shown in Fig.~\ref{fig:kaos_unscaled_v2}
scaled by the passing time.}
\label{fig:kaos_scaled_v2} 
\end{minipage}
\end{figure}

	If the passage time is long compared to 
the expansion time, spectator nucleons serve to block the path of participant hadrons emitted 
toward the reaction plane, and nuclear matter is squeezed-out perpendicular to this plane 
giving rise to negative elliptic flow. The squeeze-out contribution should then reflect
the ratio $c_s /\left( {\gamma _0 v_0 } \right)$. This is put into evidence in 
Fig.~\ref{fig:kaos_scaled_v2} where the differential elliptic flow values $v_2(p_T)$, 
shown for beam energies of 0.4 - 2 AGeV in Fig.~\ref{fig:kaos_unscaled_v2}, are scaled 
by the passing time. The rather good scaling observed suggest that $c_s$ does not change 
significantly over this beam energy range. 

For very short passage times (such as those at top AGS energies 
and beyond), the inertial confinement of participant matter is significantly reduced and preferential 
in-plane emission or positive elliptic flow is favored. This is so because the geometry of the participant region 
exposes a larger surface area in the direction of the reaction plane and pressure gradients in this 
direction are also larger. Thus, the observed trends ( ie. negative for beam energies $< 4$~AGeV 
and positive for beam energies $ > 4$~AGeV) for the beam energy dependence of elliptic flow 
are well understood \cite{sor97,dan98,pin99}. For RHIC energies, strong Lorentz contraction and very 
short passage times lead to significant reduction of shadowing effects, and positive elliptic flow is 
driven primarily by the anisotropies in the transverse 
pressure gradients \cite{olli92,teaney2001,kolb2001}.

\begin{figure}[thb]
\begin{minipage}[t]{0.5\linewidth}
\includegraphics[width=1.\linewidth]{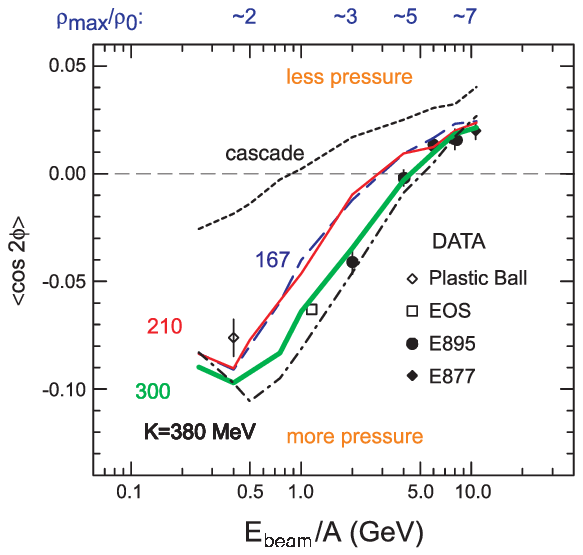}
\vskip -1.0cm
\caption{ Differential flow $v_2$ vs $E_{beam}$ for the 
energy range 0.4~$<E_{beam}<$~12.0 AGeV. Results are shown for 
protons for mid-central collisions \cite{Danielewicz:2002pu}.}  
\label{fig:v2_eos}
\end{minipage}
\hskip 0.2cm 
\begin{minipage}[t]{0.5\linewidth}
\includegraphics[width=1.\linewidth]{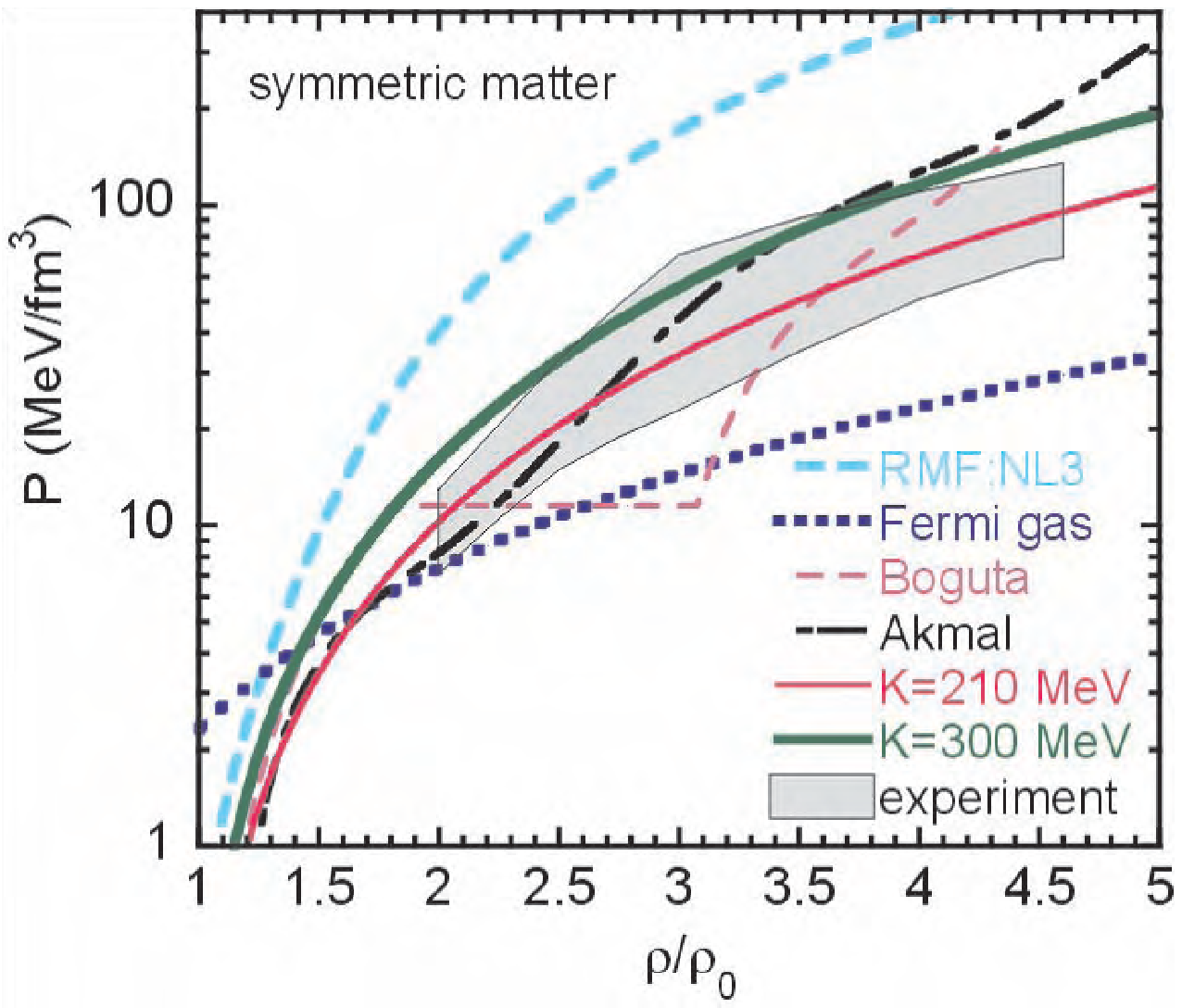}
\vskip -1.0cm
\caption{ Zero temperature EOS. The shaded region
is the region of pressures consistent with the experimental flow data.
The curves show predictions for different EOS's \cite{Danielewicz:2002pu}.}
\label{fig:eos} 
\end{minipage}
\end{figure}

\subsection{Pre-RHIC constraints for the EOS }

The elliptic and directed flow observables are sensitive to the mean field and
to the equation of state. Consequently, comparisons of these two observables to 
model calculations can provide important constraints for the EOS.
Such comparisons have been carried out for the beam energy range 
0.4~$<E_{beam}<$~12.0 AGeV \cite{Danielewicz:2002pu}.

Figs.~\ref{fig:side_flow_systematics} and \ref{fig:v2_eos} show comparisons for 
several incompressibility constants ($K$). At incident energies of 2-6 GeV/A, for example, the
transverse flow data lie near or somewhat below (to the low pressure side of) the 
$K=210$~MeV calculations, while the elliptic flow data lie closer to the $K=300$~MeV 
calculations. The calculations without a mean field (cascade) or with a weakly repulsive 
mean field ($K=167$~MeV) provide too little pressure to reproduce either flow observable 
at higher incident energies (and correspondingly higher densities). 
The calculations with $K=167$~MeV and K=380 MeV
provide lower and upper bounds on the pressure in the density 
range $2.0<\rho/\rho_{0} <5.0$.

	If one factors in the uncertainties due to the momentum dependencies of the mean fields and the 
collision integral, a range of pressure-density relationships can be established
from the comparisons made in Figs.~\ref{fig:side_flow_systematics} and \ref{fig:v2_eos}. 
These bounds on the equation of state are shown for zero temperature matter by the shaded 
region in Fig.~\ref{fig:eos}. They constitute the most current set of constraints on the EOS 
for high energy density nuclear matter created at high baryon densities, and can provide a 
rudimentary base-line for comparisons involving the essentially baryon-free 
matter created at RHIC.
\section{Elliptic flow at RHIC}
\subsection{The nature of the ``soft" matter formed at RHIC}
\begin{figure}[thb]
\begin{minipage}[t]{0.5\linewidth}
\includegraphics[width=1.\linewidth]{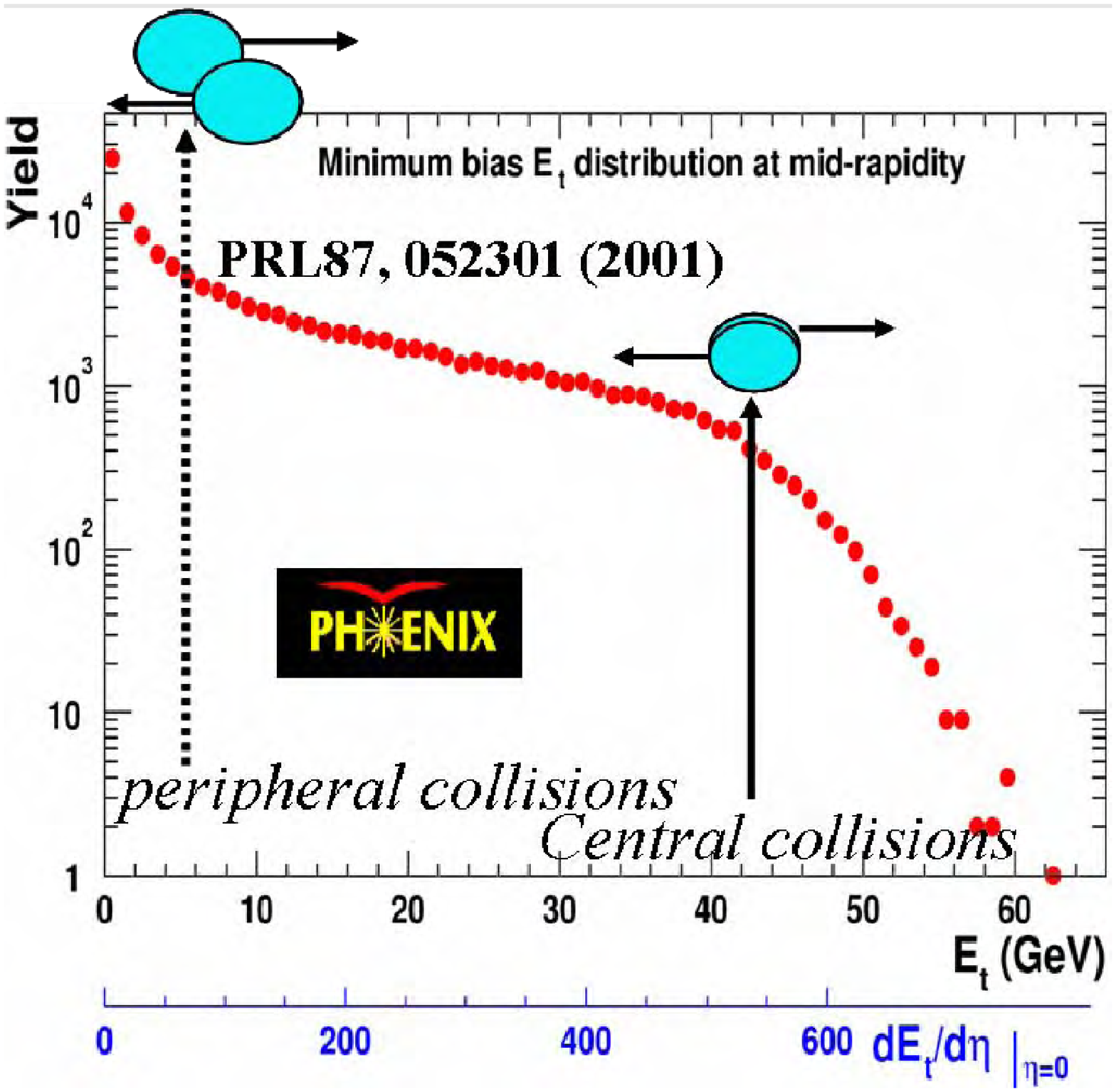}
\vskip -0.8cm
\caption{ Yield vs transverse energy $E_t$. The data are 
obtained from Ref.~\cite{phenix_et}.}  
\label{fig:et_dist_phenix}
\end{minipage}
\hskip 0.2cm 
\begin{minipage}[t]{0.5\linewidth}
\includegraphics[width=1.\linewidth]{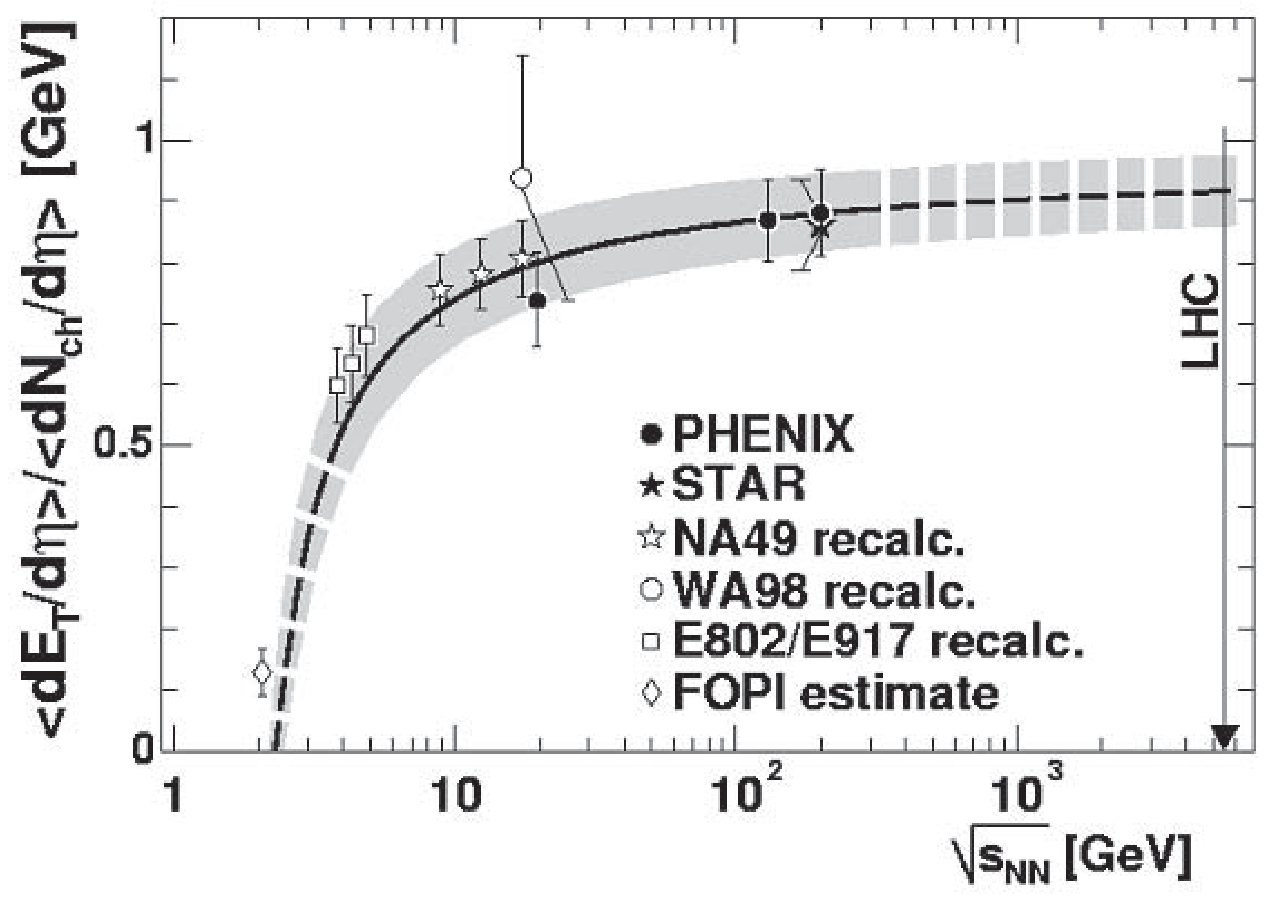}
\vskip -0.8cm
\caption{ $\left\langle E_T\right\rangle/\left\langle N_{ch}\right\rangle$ vs. 
$\sqrt{s_{NN}}$ for Au+Au collisions. The data are taken from Ref.~\cite{phenix_et_nchg}.}
\label{fig:et_ncharge_phenix} 
\end{minipage}
\vskip -0.6cm
\end{figure}
%
%
It is a well known fact that rather high energy densities are achieved in 
central and semi-central heavy ion collisions at RHIC \cite{phenix_et,phobos_mult}.
Fig.~\ref{fig:et_dist_phenix} shows a recent measurement of the transverse energy $E_T$, 
distribution for central Au+Au collisions at $\sqrt{s_{NN}} = 130$ GeV. Following the 
commonly exploited Bjorken ansatz
\[
\varepsilon _{Bj}  = \frac{1}
{{\pi \,R^2 }}\;\frac{1}
{{\tau _0 }}\;\frac{{dE_T }}
{{dy}},
\]
if we assume a thermalization time $\tau_0 \sim 0.2 - 1$~fm/c, one obtains the energy density 
$\varepsilon_{Bj}= 5~-~15$~GeV/fm$^{3}$, which is 35~-~100 times 
the energy density $\varepsilon_{0}$, for normal nuclear matter. This 
energy density far exceeds the lattice QCD estimate ($\sim 1$~GeV/fm$^{3}$) for creating a 
de-confined phase of quarks and gluons (QGP)\cite{fodor-katz02,bielefeld-QCD}. 
Fig.~\ref{fig:et_ncharge_phenix} shows the $\sqrt{s_{NN}}$ dependence of 
the ratio $\left\langle E_T\right\rangle/\left\langle N_{ch}\right\rangle$, obtained 
for Au+Au collisions spanning the collision energy range SIS~-~RHIC. It shows that the 
$\left\langle E_T\right\rangle$ rises faster than the mean multiplicity for charge particle 
production $\left\langle N_{ch}\right\rangle$, from SIS 
to SPS energies. However, there is little change in the  
ratio $\left\langle E_T\right\rangle/\left\langle N_{ch}\right\rangle$
from SPS to RHIC, suggesting that the $\sqrt{s_{NN}}$ increase results primarily in an 
increase in particle production. 
\begin{figure}[!htb]
\begin{minipage}[t]{0.5\linewidth}
\includegraphics[width=1.\linewidth]{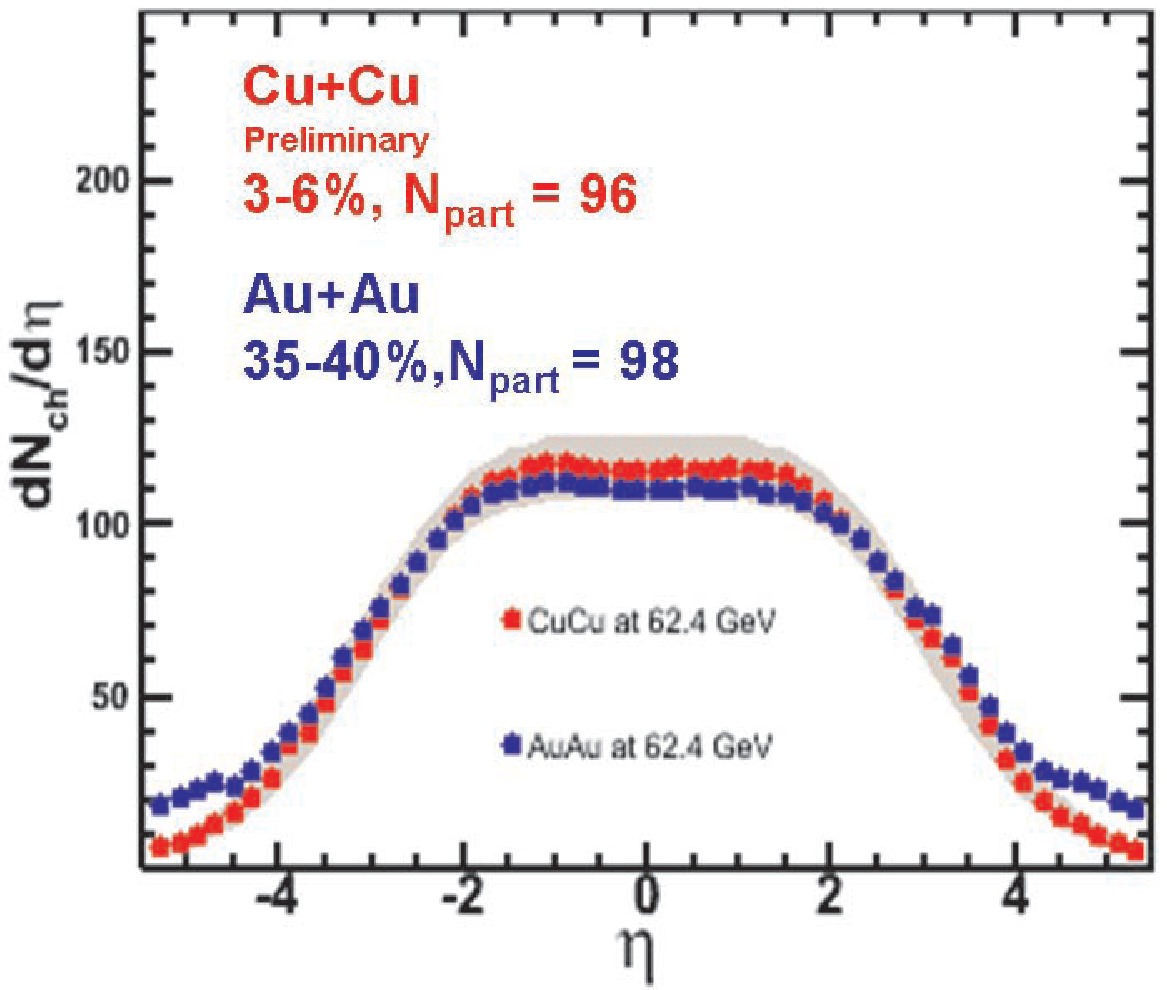}
\vskip -1.2cm
\caption{ Comparison of the multiplicity distributions for Cu+Cu and 
Au+Au collisions obtained for the same number of participants at 
$\sqrt{s_{NN}} = 62.4$ GeV. The data are obtained from Ref.~\cite{phobos_mult}.}  
\label{fig:phobos_dndeta63}
\end{minipage}
\hskip 0.2cm 
\begin{minipage}[t]{0.5\linewidth}
\includegraphics[width=1.\linewidth]{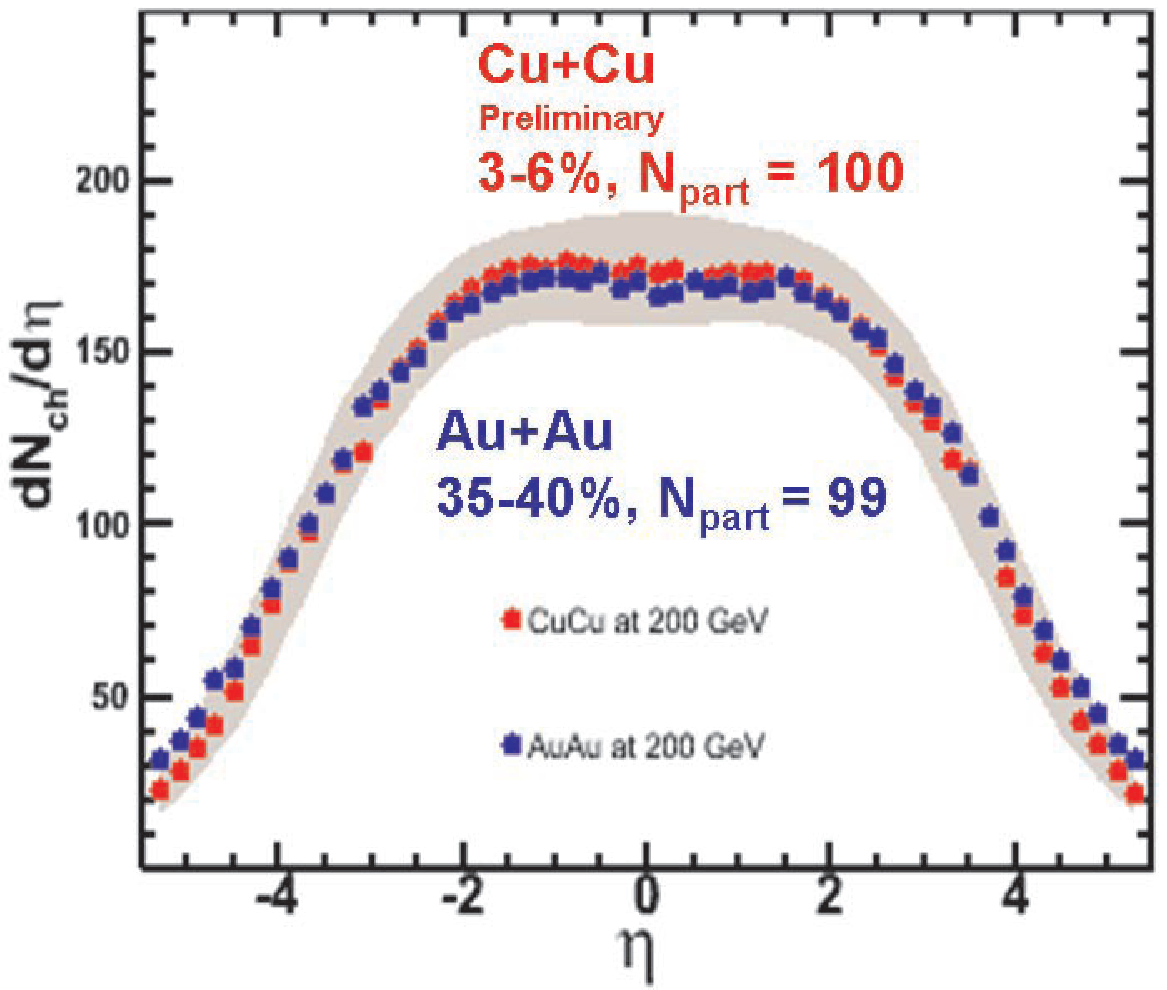}
\vskip -1.2cm
\caption{ Comparison of the multiplicity distributions for Cu+Cu and 
Au+Au collisions obtained for the same number of participants at 
$\sqrt{s_{NN}} = 200.0$ GeV. The data are obtained from Ref.~\cite{phobos_mult}.}
\label{fig:phobos_dndeta200} 
\end{minipage}
\vskip -0.6cm
\end{figure}

There are strong hints that this particle production follows rapid thermalization. 
First, the multiplicity distribution for the same number of participants is independent 
of colliding system size, as would be expected from a system which ``forgets"  how it is 
formed. A beautiful demonstration of this is shown in Figs.~\ref{fig:phobos_dndeta63}
and \ref{fig:phobos_dndeta200} with PHOBOS data for Au+Au and Cu+Cu collisions at two 
separate collision energies. 
\begin{figure}[!htb]
\begin{center}
\begin{minipage}[t]{0.8\linewidth}
\includegraphics[width=1.0\linewidth]{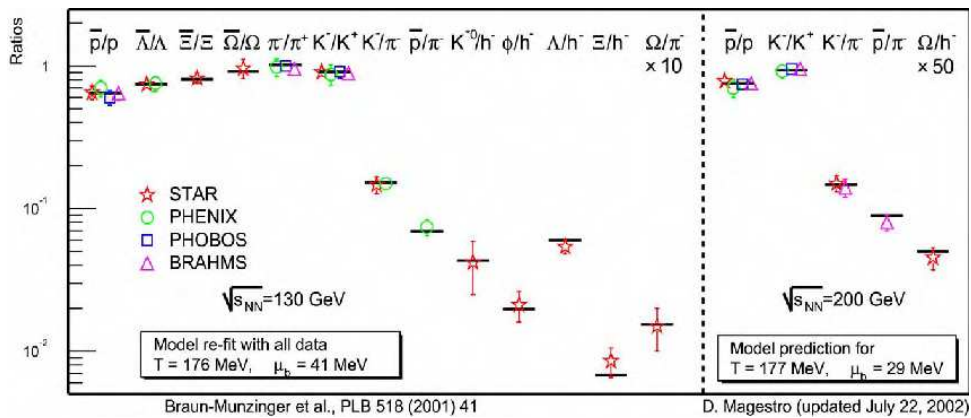}
\vskip -1.0cm
\caption{ Comparison of measured particle ratios for Au+Au collisions, with 
the predictions of a statistical model (solid lines).}
\label{fig:particle_ratios} 
\end{minipage}
\end{center}
\vskip -0.6cm
\end{figure}
Second, a comparison of the measured anti-particle to particle ratios with 
the predictions of a statistical model
\[
\frac{{\bar p}}
{p} = \frac{{e^{ - (E + \mu )/T} }}
{{e^{ - (E - \mu )/T} }} = e^{ - 2\mu /T}, 
\]
gives excellent agreement as shown for Au+Au collisions in 
Fig.~\ref{fig:particle_ratios}. It is noteworthy that the the temperature $T \sim 176$~MeV, 
required for this excellent agreement compares well with the critical temperature 
$T_{cr}\sim 170$~MeV, required for the QGP phase transition. 
The baryochemical potential $\mu$, extracted for this temperature, increases 
from $\sim 29$~MeV at $\sqrt{s_{NN}} = 200.0$ GeV to $\sim 40$~MeV 
at $\sqrt{s_{NN}} = 130.0$ GeV.

Given the large energy densities produced in RHIC collisions, if thermalization does indeed 
occur, then one expects the development of large pressures;
\[
{{P} = \rho \cdot \left. {\left( {{\raise0.7ex\hbox{${\partial \varepsilon }$} \!\mathord{\left/
 {\vphantom {{\partial \varepsilon } {\partial \rho }}}\right.\kern-\nulldelimiterspace}
\!\lower0.7ex\hbox{${\partial \rho }$}}} \right)} \right|_{{{s/}}\rho } }.
\]
More importantly, if thermalization is rapid in non-central collisions, then large 
pressure gradients resulting from the initial spatial anisotropy or 
eccentricity 
\[
\epsilon = \left( {\left\langle {y^2 } \right\rangle  - \left\langle {x^2 } 
\right\rangle } \right)/\left( {\left\langle {y^2 } \right\rangle  + 
\left\langle {x^2 } \right\rangle } \right),
\]
of the collision zone would lead to strong elliptic flow. That is, the observation of 
large elliptic flow for a variety of particle species would be compatible with the 
expectation for early thermalization. 

\subsection{Elliptic flow results}
\begin{figure}[!htb]
\begin{minipage}[t]{0.5\linewidth}
\includegraphics[width=1.\linewidth]{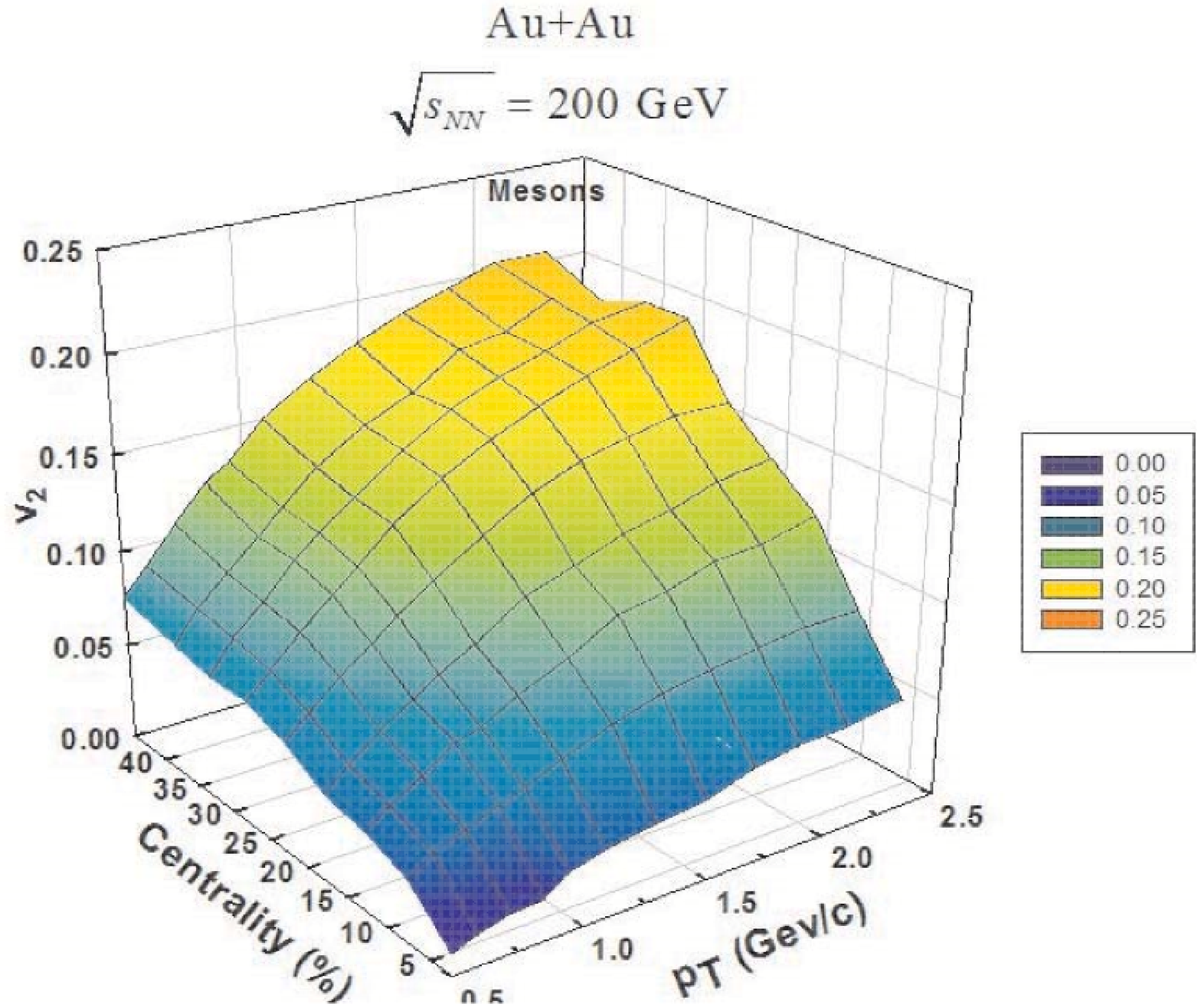}
\vskip -0.8cm
\caption{ PHENIX preliminary data for the doubly differential 
anisotropy $v_2(centrality, p_T)$, for mesons (dominantly pions).}  
\label{fig:mesons_2d}
\end{minipage}
\hskip 0.2cm 
\begin{minipage}[t]{0.5\linewidth}
\includegraphics[width=1.\linewidth]{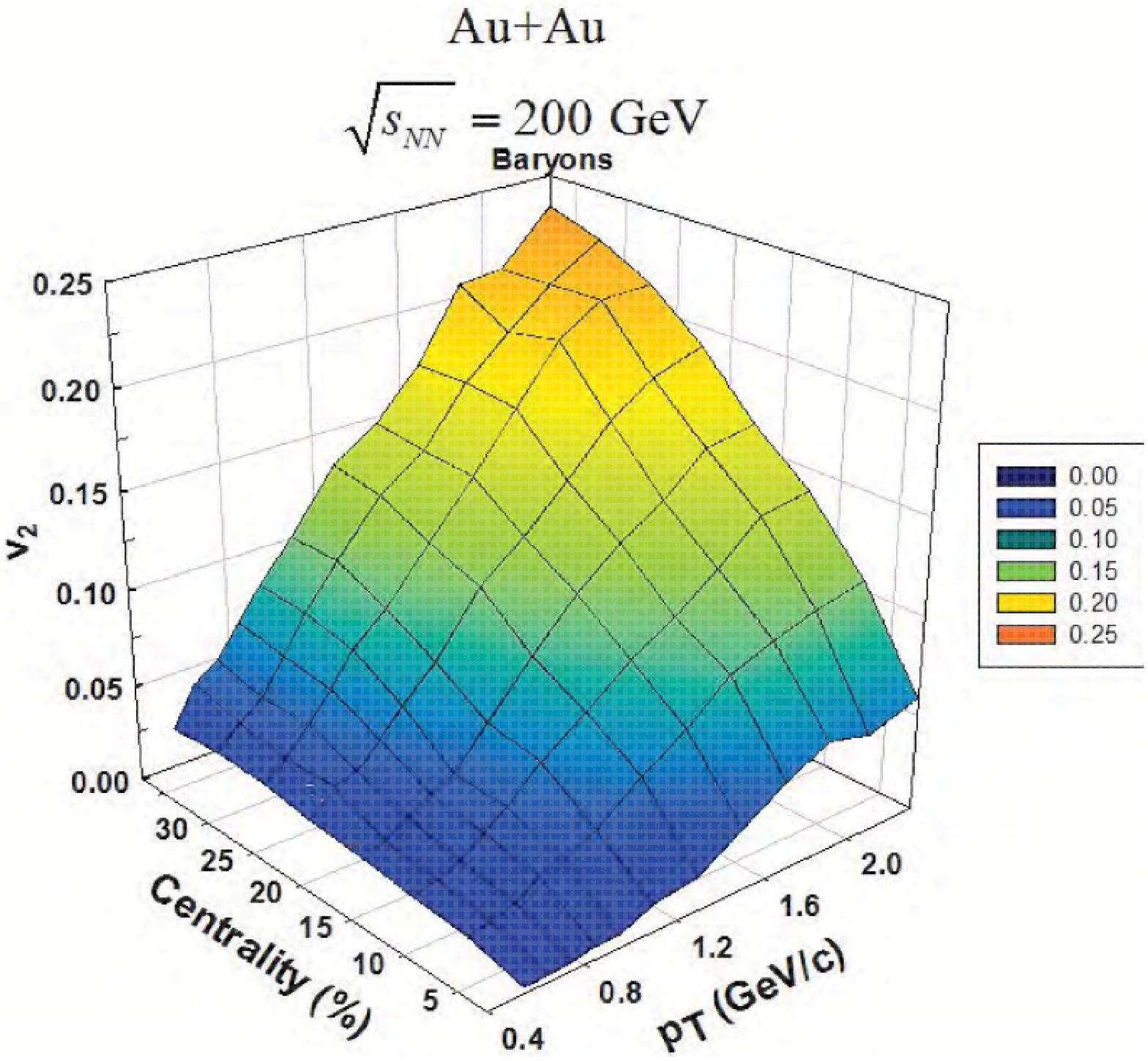}
\vskip -0.8cm
\caption{ PHENIX preliminary data for the  doubly differential 
anisotropy $v_2(centrality, p_T)$, for protons and anti-protons.}
\label{fig:baryons_2d} 
\end{minipage}
\vskip -0.6cm
\end{figure}
	Detailed differential and integral measurements are now available for charged hadrons and 
a variety identified particle species 
($\mathrm{p},\pi ,K,\Lambda ,\Omega ,\phi ,\Xi ,\mathrm{d},\mathrm{D}$) 
at RHIC \cite{qm05_flow}. 
Figs.~\ref{fig:mesons_2d} and~\ref{fig:baryons_2d} summarize the extracted 
doubly differential anisotropy $v_2(centrality, p_T)$, for mesons (dominantly pions) 
and baryons (protons and anti-protons). They give an excellent overview of the 
the detailed evolution of $v_2$ as centrality and $p_T$ are varied. The results shown 
for protons ($p$) and anti-protons ($\overline{p}$) give an especially good view of 
the evolution away from the well known quadratic dependence of $v_2(p_T)$ 
(which is also observed in very central collisions for these data) as the 
collisions become more peripheral. 
\begin{figure}[!htb]
\begin{minipage}[t]{0.5\linewidth}
\includegraphics[width=1.\linewidth]{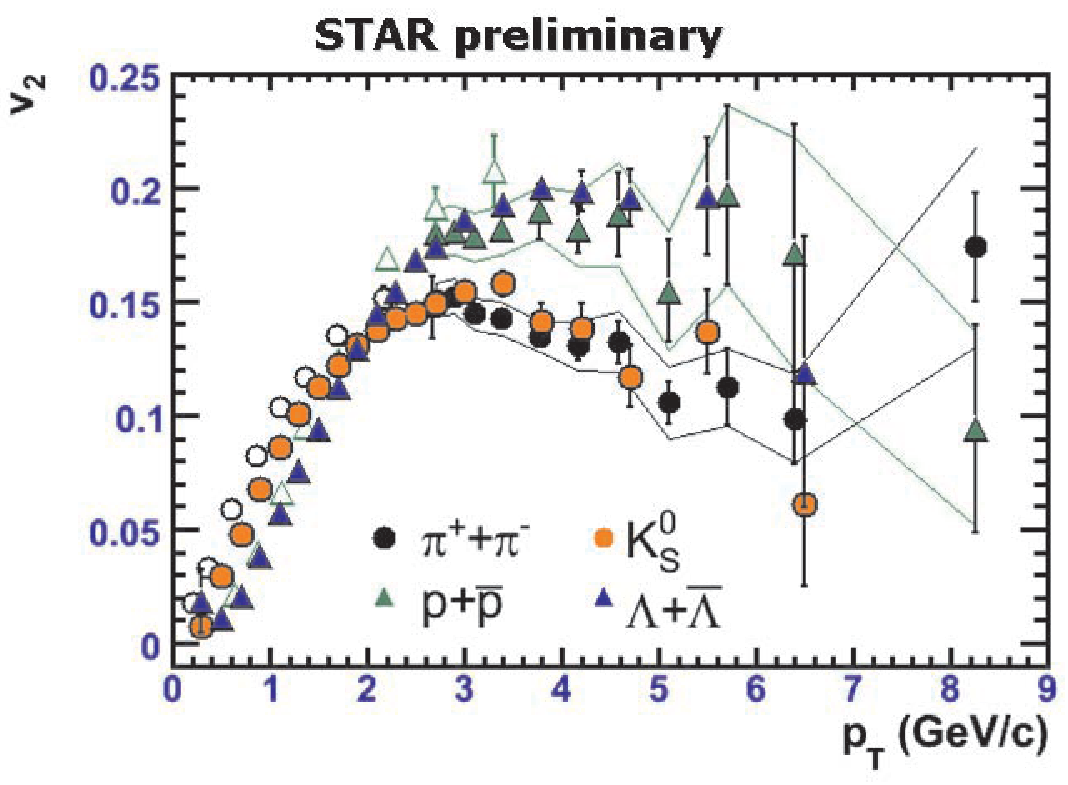}
\vskip -1.2cm
\caption{ Preliminary STAR data showing $v_2(p_T)$ vs. $p_T$ for different 
particle species produced in Au+Au collisions at $\sqrt{s_{NN}} = 200$~GeV
\cite{qm05_flow}.}  
\label{fig:pid_v2_star}
\end{minipage}
\hskip 0.2cm 
\begin{minipage}[t]{0.5\linewidth}
\includegraphics[width=1.\linewidth]{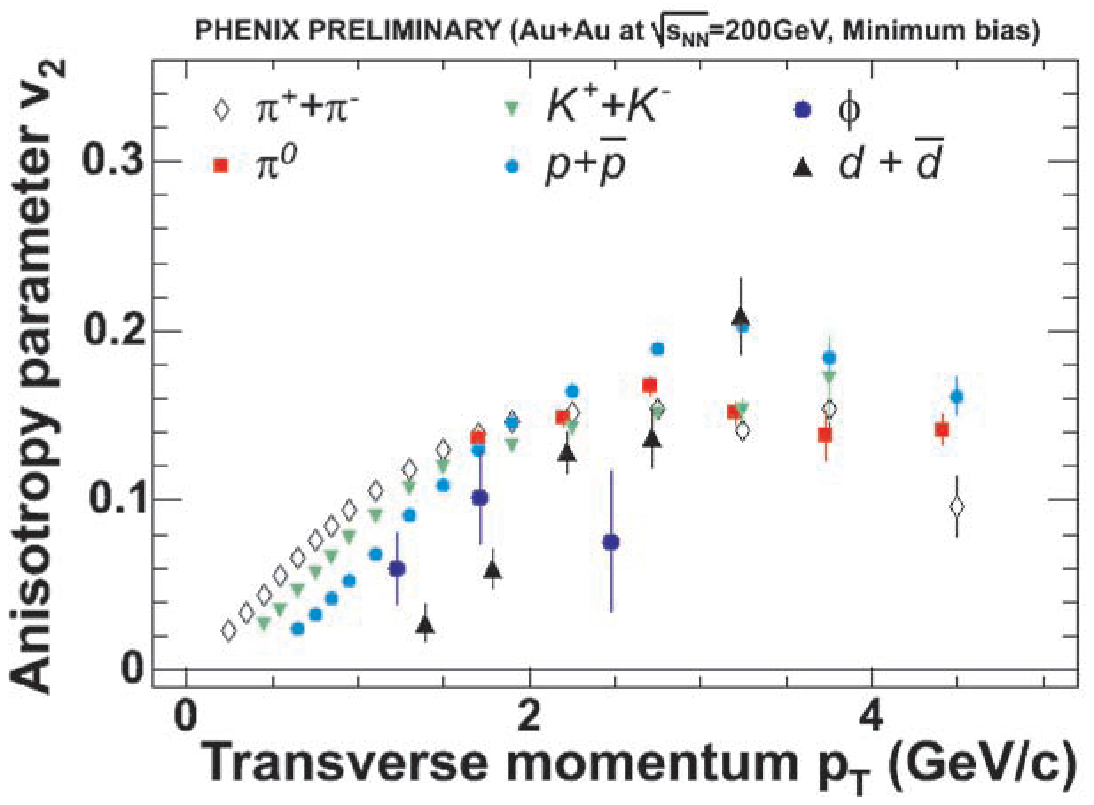}
\vskip -1.2cm
\caption{ Preliminary PHENIX data showing $v_2(p_T)$ vs. $p_T$ for different 
particle species produced in Au+Au collisions at $\sqrt{s_{NN}} = 200$~GeV
\cite{qm05_flow}.}
\label{fig:pid_v2_phenix} 
\end{minipage}
\vskip -0.6cm
\end{figure}
Figs.~\ref{fig:pid_v2_star} and \ref{fig:pid_v2_phenix}, show a representative 
set of differential flow measurements $v_2(p_T)$ for several different particle 
species. They are all characterized by rather large magnitudes compatible with 
the predictions of the hydrodynamic model \cite{teaney2001,kolb2001,hirano_qm05}, 
which in turn implies the creation of a strongly interacting medium 
and essentially full local thermal equilibrium. It is especially noteworthy 
that the $v_2$ values for the $\phi$ are comparable to those of other hadrons.
Given its rather small re-scattering cross section, such $v_2$ values suggest 
that thermalization is rapid and pressure gradients develop at the 
partonic level.  

\begin{figure}[!htb]
\begin{minipage}[t]{0.5\linewidth}
\includegraphics[width=1.\linewidth]{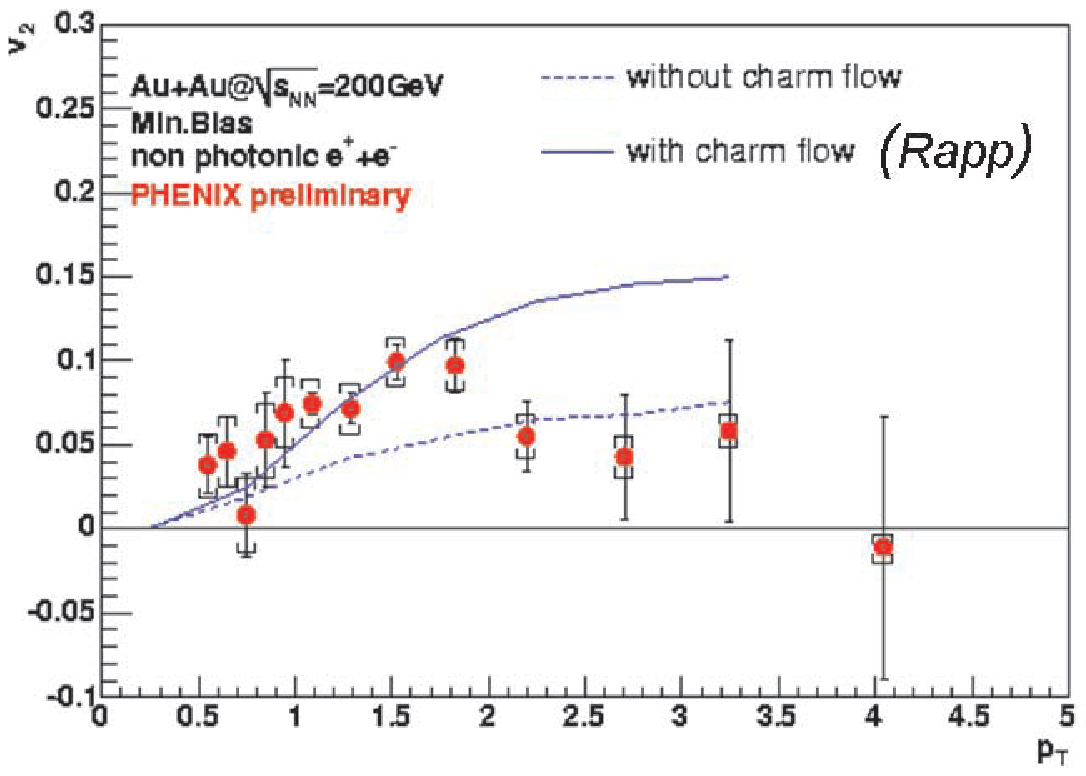}
\vskip -1.2cm
\caption{ $v_2(p_T)$ vs. $p_T$ for charm \cite{butsyk_QM05}. 
The lines compare the results from a calculation which 
includes/excludes charm flow \cite{rapp_qm05}.}  
\label{fig:charm_flow}
\end{minipage}
\hskip 0.2cm 
\begin{minipage}[t]{0.5\linewidth}
\includegraphics[width=1.\linewidth]{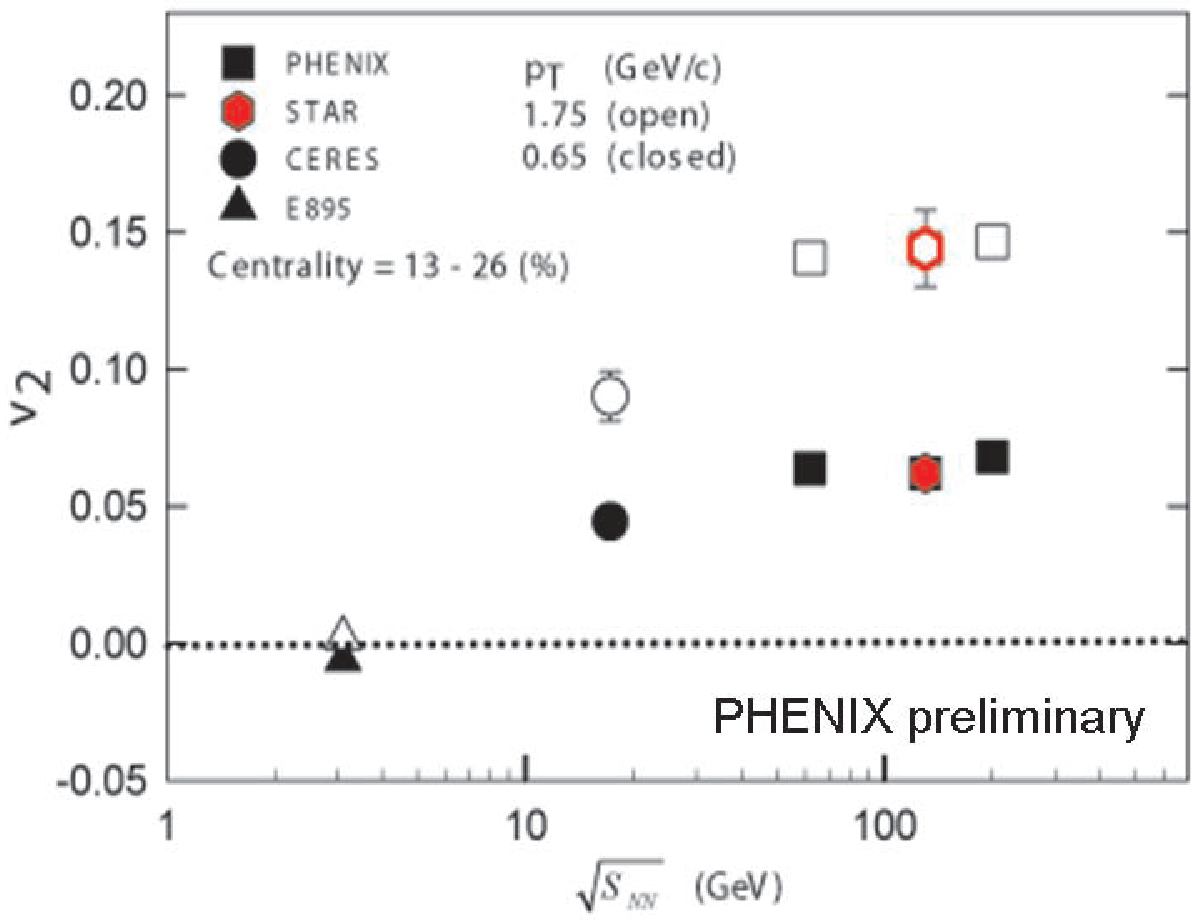}
\vskip -1.2cm
\caption{ $\sqrt{s_{NN}}$ dependence of $v_2$ for fixed $p_T$  selections of 
0.65 and 1.75~GeV/c from A+A collisions with A~$\approx 200$ \cite{v2_roots}.}
\label{fig:v2_roots} 
\end{minipage}
\vskip -0.6cm
\end{figure}
Further evidence for rapid thermalzation can also be found in the 
recent measurements of $v_2$ for non-photonic electrons \cite{butsyk_QM05}. 
Such measurements provide a probe for whether or not the charm quark flows. 
Fig.~\ref{fig:charm_flow} compares recent PHENIX measurements to the calculations 
of Rapp et al. \cite{rapp_qm05}. The good agreement (for $p_T < 2$~GeV/c) between 
the data and the calculation which includes charm flow, suggests heavy quark 
thermalization in RHIC collisions. 

	As discussed earlier, the pressure or energy density gradients provide the fluid 
acceleration necessary to develop the large observed values of $v_2$. 
The initial energy density controls how much flow can develop globally, since it sets 
the overall time scale between the beginning of hydrodynamic expansion and final decoupling.
On the other hand, the detailed development of the flow patterns are 
controlled by the temperature dependent speed of sound $c_s^2  = \partial P/\partial \varepsilon$.
Lattice QCD calculations indicate that $c_s^2 \approx 1/3$ for $T > 2T_{cr}$, but drops steeply 
by more than a factor of six near the ``softest point" where $T \approx T_{cr}$.
Thereafter, it rises again in the hadron resonance gas phase to the value 
$c_s^2 \approx 0.15$. Given this, an important question is whether or not there 
might be a trace of this ``softest point" in the data.

Fig.~\ref{fig:v2_roots} shows the $\sqrt{s_{NN}}$ dependence of $v_2$ over a broad 
range of bombarding energies for two $p_T$  selections. The data show essentially no beam 
energy dependence over the entire energy range explored at RHIC, but decreases substantially 
as one moves down to the lower SPS and AGS energies. These data reflect the
predicted non-monotonic structure in the elliptic flow excitation function \cite{kolb_PRC62}.
However, they do not show the decrease (due to very strong softening of the EOS in the phase
transition region) followed by a recovery in the moderately stiff hadron 
gas phase predicted by the model \cite{kolb_PRC62}. Nonetheless, the apparent saturation 
of $v_2$ above $\sqrt{s_{NN}}$~=~62.4 GeV does not exclude the role of a rather soft 
equation of state resulting from the production of a mixed phase~\cite{PHENIX_ppg47} 
for the range $\sqrt{s_{NN}}$~=~62.4~-~200 GeV.

\section{Universal scaling and perfect fluid hydrodynamics}
A particularly important question of great 
current interest is whether or not the fine structure of azimuthal anisotropy 
(ie. its detailed dependence on centrality, transverse momentum, 
particle type, higher harmonics, etc) can provide valuable constraints on 
(i) the range of validity of perfect fluid hydrodynamics, (ii) the parameters of the model
and (iii) the onset of competing mechanisms such as 
quark-coalescence~\cite{Fries:2003vb}. One approach to address such questions, 
is the detailed investigation of flow data to test the validity and/or failure of several 
scaling ``laws" predicted by perfect fluid hydrodynamics. 

 An important scaling prediction of hydrodynamic theory is exemplified by 
the exact analytic hydro solutions~\cite{Csorgo:2001xm} exploited in the Buda-Lund 
model~\cite{Csorgo:1995bi,Csanad:2003qa}.
The model gives:
\begin{equation}
v_{2n}=\frac{I_n(w)}{I_0(w)}, n=1,2,.., {\mathrm{  }}w = \frac{p_t^2}{4
   \overline{m}_t} \left(\frac{1}{T_{y}}
   -\frac{1}{T_{x}}\right),
\label{eq2}
\end{equation}
where $I_{0,n}$ are modified Bessel-functions, 
$\overline{m}_t$ is a rapidity dependent average transverse mass                    
(at midrapidity, $\overline{m_T} = m_T$, see ref.~\cite{Csanad:2003qa} for details), 
and $T_{x}$ and  $T_{y}$ are direction ($x$ and $y$) dependent slope parameters:
\begin{eqnarray}
   T_{x}&=&T_0+\overline{m}_t \, \dot X_f^2
       \frac{T_0}{T_0 +\overline{m}_t a},\\
   T_{y}&=&T_0+\overline{m}_t \, \dot Y_f^2
     \frac{T_0}{T_0 +\overline{m}_t a}.
\label{eq3}
\end{eqnarray}
Here, $\dot X_f$ and $\dot Y_f$ gives the transverse expansion rate of the 
fireball at freeze-out, and $a = (T_0-T_s)/T_s$ is its transverse 
temperature inhomogeneity, characterized by the temperature at its    
center $T_0$, and at its surface $T_s$. The thermal and collective contributions
can be made more transparent by replacing $p_T$ with the transverse
rapidity $y_T=0.5\log((m_T+p_T)/(m_T-p_T))$, to give 
the following approximate scaling law:
\begin{eqnarray}
v_2 \sim \frac{{k_{1} }}
{{T_0 }} \times y_T^2 m\left( {1 + \frac{{k_{2} }}
{{k_{1} }}\frac{{T_0 }}
{m} + \frac{{k_3}}
{{k_1 }}\left( {\frac{{T_0 }}
{m}} \right)^2  + ..} \right),
\label{eq4}
\end{eqnarray}
where $k_{1,2,3}$ are mass ($m$) dependent constants. 
Equation~\ref{eq4} shows that $v_2$ should scale 
as $\sim k_m \times m y_T^2$
for different particle species or flavor. Hereafter, 
we define the scaled variable $y^{fs}_T \equiv k_m \times m y_T^2$.
The higher harmonic $v_4 \sim \frac{1}{2}v_2^2 + k_4 y_T^4$, 
can be similarly obtained from Eq.~\ref{eq2}. It is also easy to 
show that $v_2$ scales with eccentricity and should be independent 
of the mass (size) of the colliding system for the same eccentricity. 
In what follows, we test whether or not such scaling is evidenced by RHIC data.
The extent to which such scaling holds, gives an indication of the applicability
and validity of perfect fluid hydrodynamics.

\subsection{ Eccentricity scaling \& system size dependence}
\begin{figure}[thb]
\begin{minipage}[t]{0.5\linewidth}
\includegraphics[width=1.\linewidth]{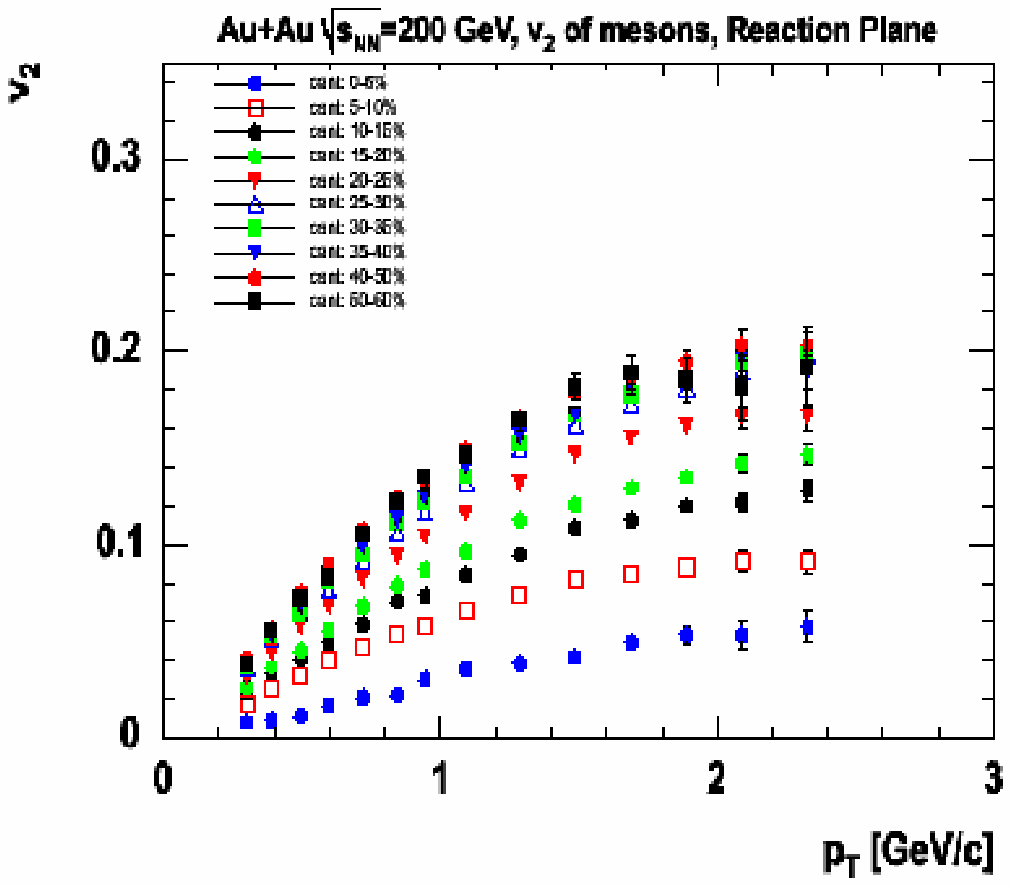}
\vskip -0.8cm
\caption{ Doubly differential anisotropy $v_2(centrality,p_T)$ 
vs. $p_T$. Preliminary PHENIX data are shown for several centrality 
selections for Au+Au collisions \cite{qm05_flow}.}  
\label{fig:v2_pt_unscaled}
\end{minipage}
\hskip 0.2cm 
\begin{minipage}[t]{0.5\linewidth}
\includegraphics[width=1.\linewidth]{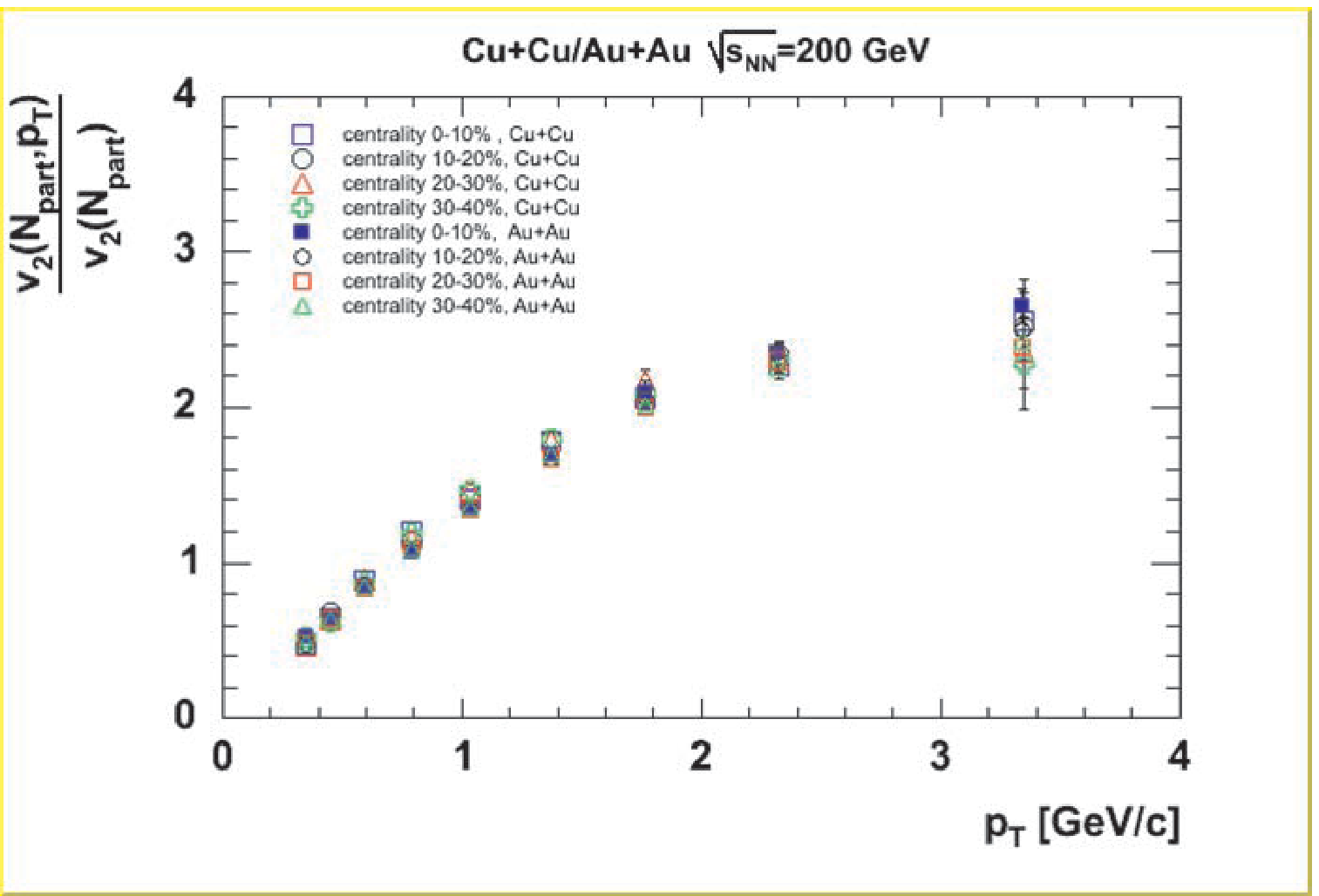}
\vskip -0.8cm
\caption{ Scaled doubly differential anisotropy $v_2(centrality,p_T)/v_2(centrality)$ 
vs. $p_T$. Preliminary PHENIX data are shown for several centrality selections for 
Au+Au and Cu+Cu collisions \cite{qm05_flow}.}
\label{fig:v2_pt_scaled} 
\end{minipage}
\vskip -0.6cm
\end{figure}
Fig.~\ref{fig:v2_pt_unscaled} show the differential $v_2(centrality,p_T)$, 
for mesons for several centralities. It shows the very well known dependence 
of $v_2$ on $p_T$ and centrality.
Fig.~\ref{fig:v2_pt_scaled} show $v_2$ results for Cu+Cu and Au+Au 
scaled by the $p_T$-integrated $v_2$ value obtained for each centrality selection. 
We note here that the $p_T$-integrated flow is monotonic and linearly related to the 
eccentricity over a broad range of centralities. The scaled data for Au+Au and 
Cu+Cu shown in Fig.~\ref{fig:v2_pt_scaled}, are clearly compatible with 
the predicted eccentricity scaling. The fact that Cu+Cu and Au+Au scale to the same value 
indicates that the scaled $v_2$ values are independent of system size as predicted. 

\subsection{ Particle flavor, pseudo-rapidity \& higher harmonic scaling}
\begin{figure}[thb]
\begin{minipage}[t]{0.5\linewidth}
\includegraphics[width=1.\linewidth]{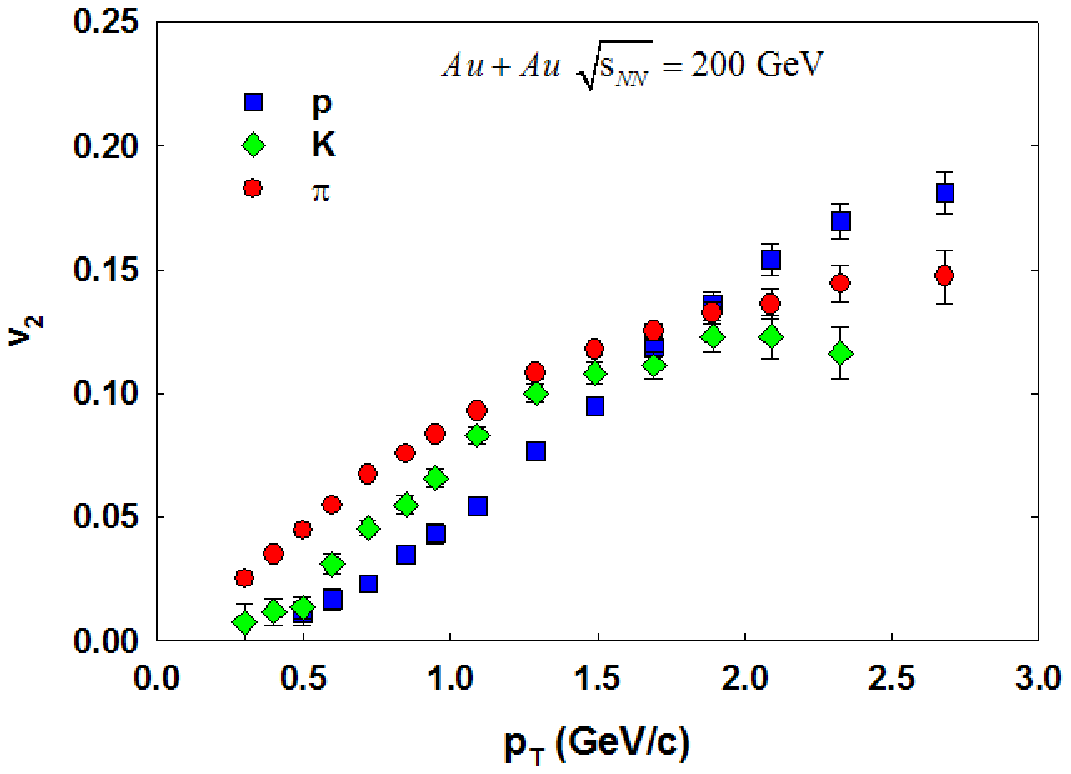}
\vskip -0.8cm
\caption{ Differential anisotropy $v_2(p_T)$ for protons, pions and 
kaons vs. $p_T$ \cite{qm05_flow}.}  
\label{fig:v2_pi-k-p_unscaled}
\end{minipage}
\hskip 0.2cm 
\begin{minipage}[t]{0.5\linewidth}
\includegraphics[width=1.\linewidth]{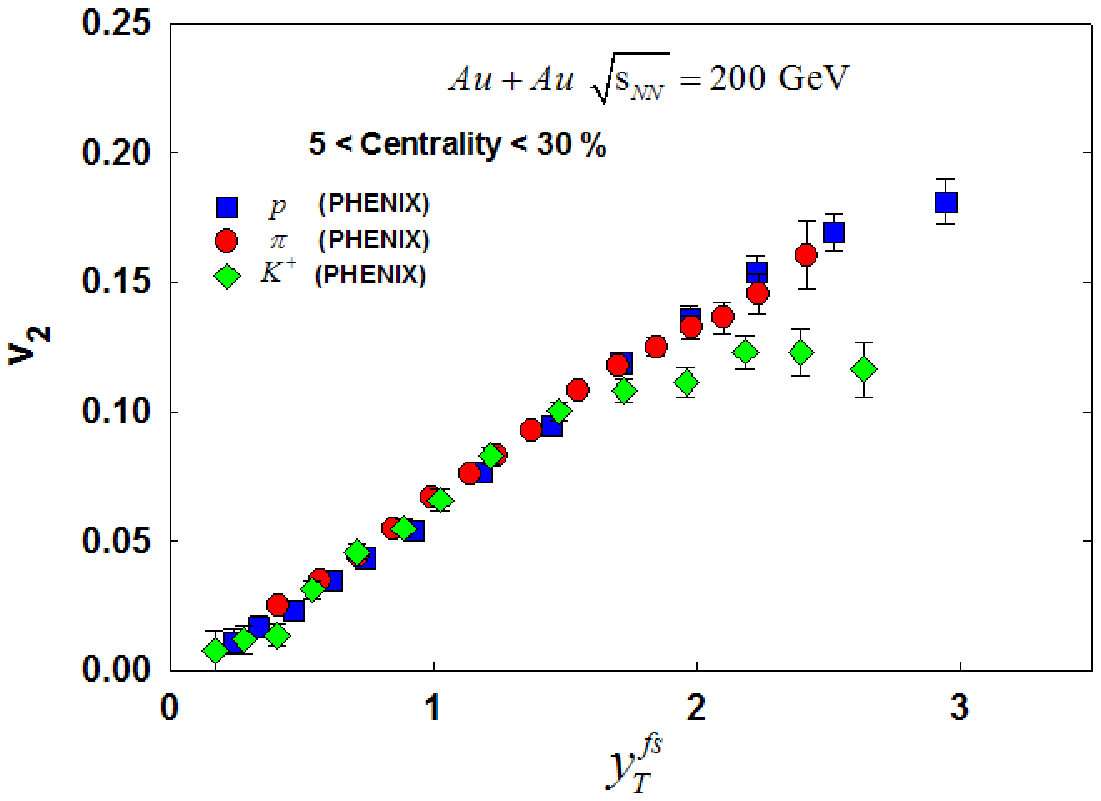}
\vskip -0.8cm
\caption{ Differential anisotropy $v_2(y^{fs}_T)$ for protons, pions and 
kaons vs. $y^{fs}_T$, see text.}
\label{fig:v2_pi-k-p_scaled} 
\end{minipage}
\vskip -0.6cm
\end{figure}
Fig.~\ref{fig:v2_pi-k-p_unscaled} shows a comparison of 
the differential anisotropy $v_2(p_T)$ for protons, kaons and pions obtained for 
the centrality selection 5-30\%. The well known and rather characteristic flavor 
dependence of $v_2$ is clearly exhibited in the figure. If this aspect of the 
fine structure of $v_2$ has a hydrodynamic origin then the prediction is that 
it should scale with the variable $y^{fs}_T \equiv k_m \times m y_T^2$.
Fig.~\ref{fig:v2_pi-k-p_scaled} shows $v_2$ data for the same particle species 
scaled by the fine structure scaling variable $y^{fs}_T$. The results indicate
scaling over a relatively broad range in $y^{fs}_T$. It is interesting to note 
here that scaling appears to break in the $y^{fs}_T$ range where quark number 
scaling seem to work \cite{Fries:2003vb}. On the other hand, the latter is based 
on the coalescence of quarks which are flowing. Thus, hydrodynamic flow appears 
to dominate the dynamics for both low and intermediate $p_T$ particles. 
\begin{figure}[!htb]
\begin{minipage}{0.5\linewidth}
\includegraphics[width=1.\linewidth]{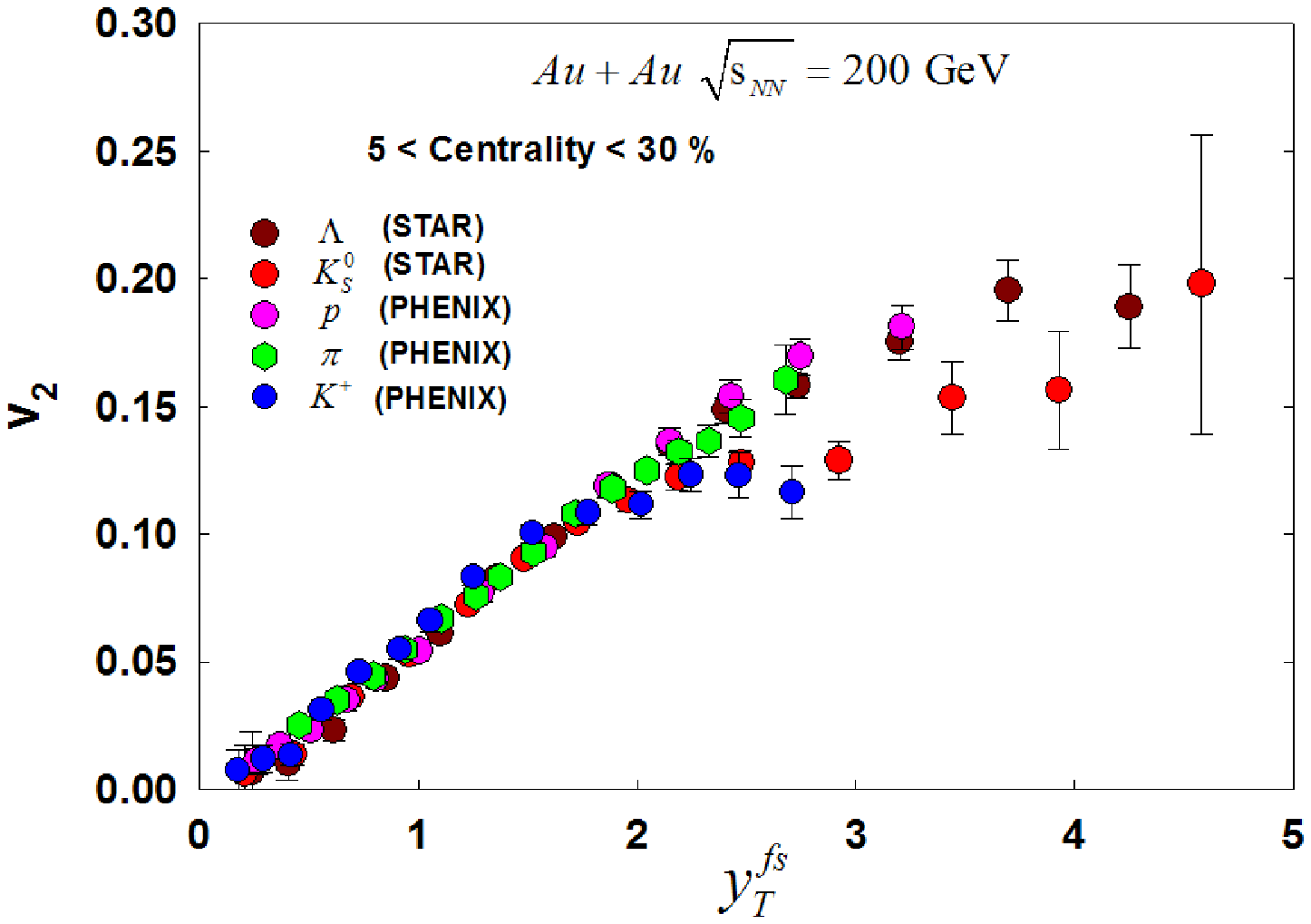}
\vskip -0.8cm
\caption{ Differential anisotropy $v_2(y^{fs}_T)$ vs. $y^{fs}_T$ for 
several different particles as indicated. The data are obtained from 
Refs. \cite{qm05_flow} and \cite{star_flow_prc}}  
\label{fig:Star_Phenix_Pid_Scaling}
\end{minipage}
%
\hskip 0.2cm 
%
\begin{minipage}{0.5\linewidth}
\includegraphics[width=1.\linewidth]{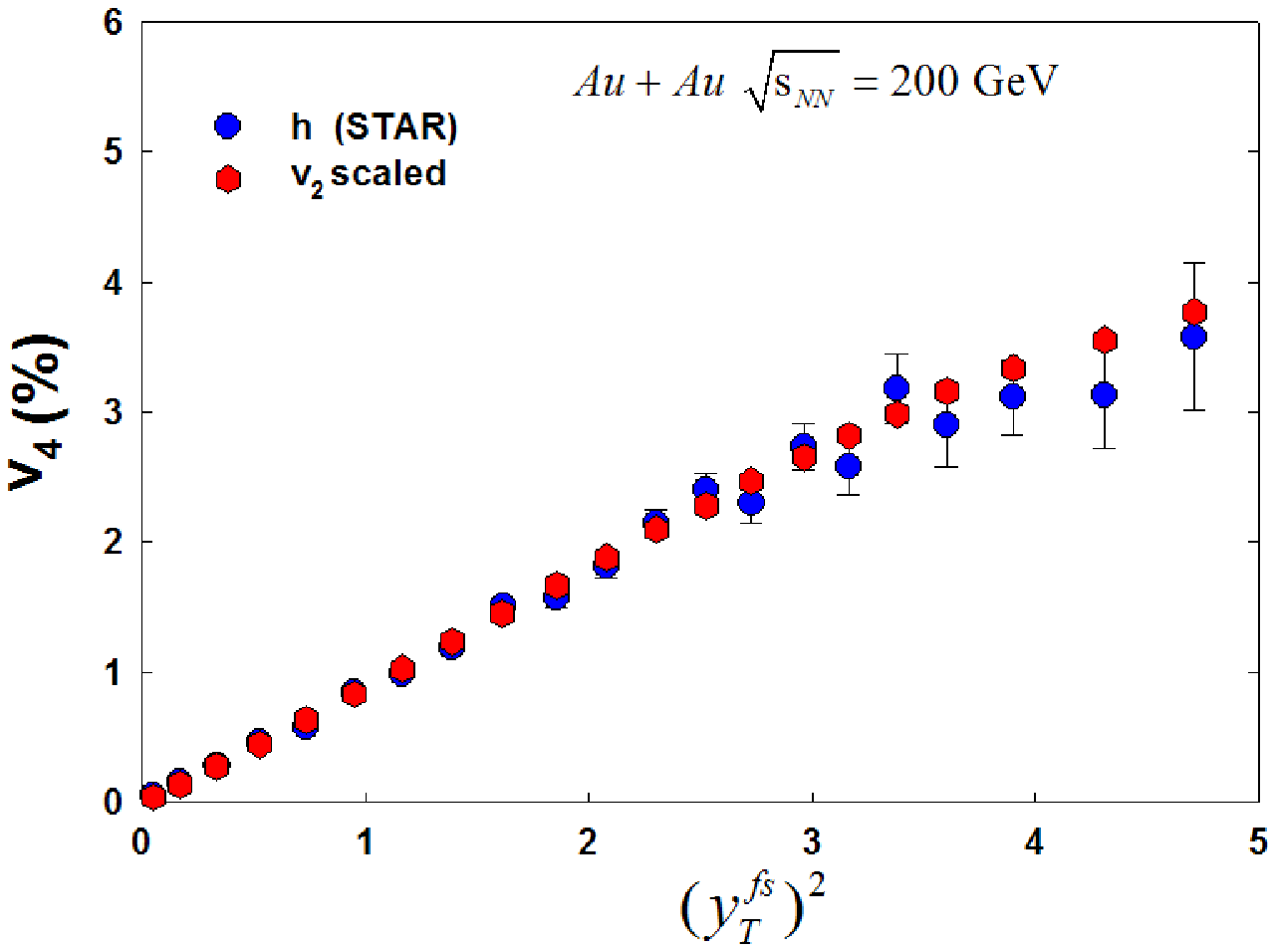}
\vskip -0.8cm
\caption{ Comparison of measured and predicted $v_4$ for 
charged hadrons. The data are obtained from Ref. \cite{star_flow_prc}.}
\label{fig:v2_v4_scaling} 
\end{minipage}
\end{figure}
Fig.~\ref{fig:Star_Phenix_Pid_Scaling} shows $v_2$ data for a much larger selection of 
particle species (from both PHENIX and STAR) scaled by the fine structure scaling 
variable $y^{fs}_T$. Here again, the results indicate scaling over a relatively 
broad range in $y^{fs}_T$. The observed flavor scaling in these 
data provides rather strong evidence that the observed anisotropy is derived from
a hydrodynamic origin. This conclusion is also supported by the good agreement 
obtained between the measured $v_4$ and the scaled 
values ($\acute{v_4} \sim \frac{1}{2}v_2^2 + k_m y_T^4$) shown in 
Fig.~\ref{fig:v2_v4_scaling}. 

	An excitation function of the pseudo-rapidity dependence of $v_2$ also serves 
as an important test for the predictions of perfect fluid 
hydrodynamics. Fig.~\ref{fig:figs/phobos_v2_eta} shows 
results from a recent analysis of PHOBOS data \cite{Csanad:2005qr}.
The theoretical curves shown in the figure, give strong indications 
that these data are consistent with the analytic predictions 
of perfect fluid hydrodynamics ~\cite{Csorgo:1995bi,Csanad:2003qa}.

\subsection{ Quark number scaling}
\begin{figure}[!htb]
\begin{minipage}{0.5\linewidth}
\includegraphics[width=1.\linewidth]{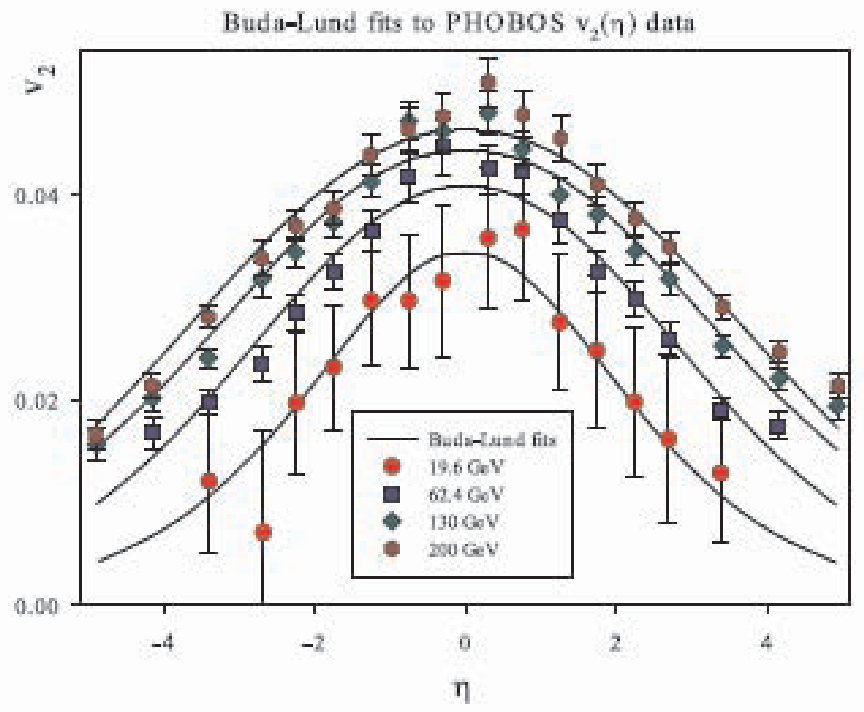}
\vskip -0.8cm
\caption{ PHOBOS data showing $v_2$ vs. $\eta$. The curves 
show the results of theoretical calculations \cite{Csanad:2005qr}.}  
\label{fig:figs/phobos_v2_eta}
\end{minipage}
\hskip 0.2cm 
\begin{minipage}{0.5\linewidth}
\includegraphics[width=1.\linewidth]{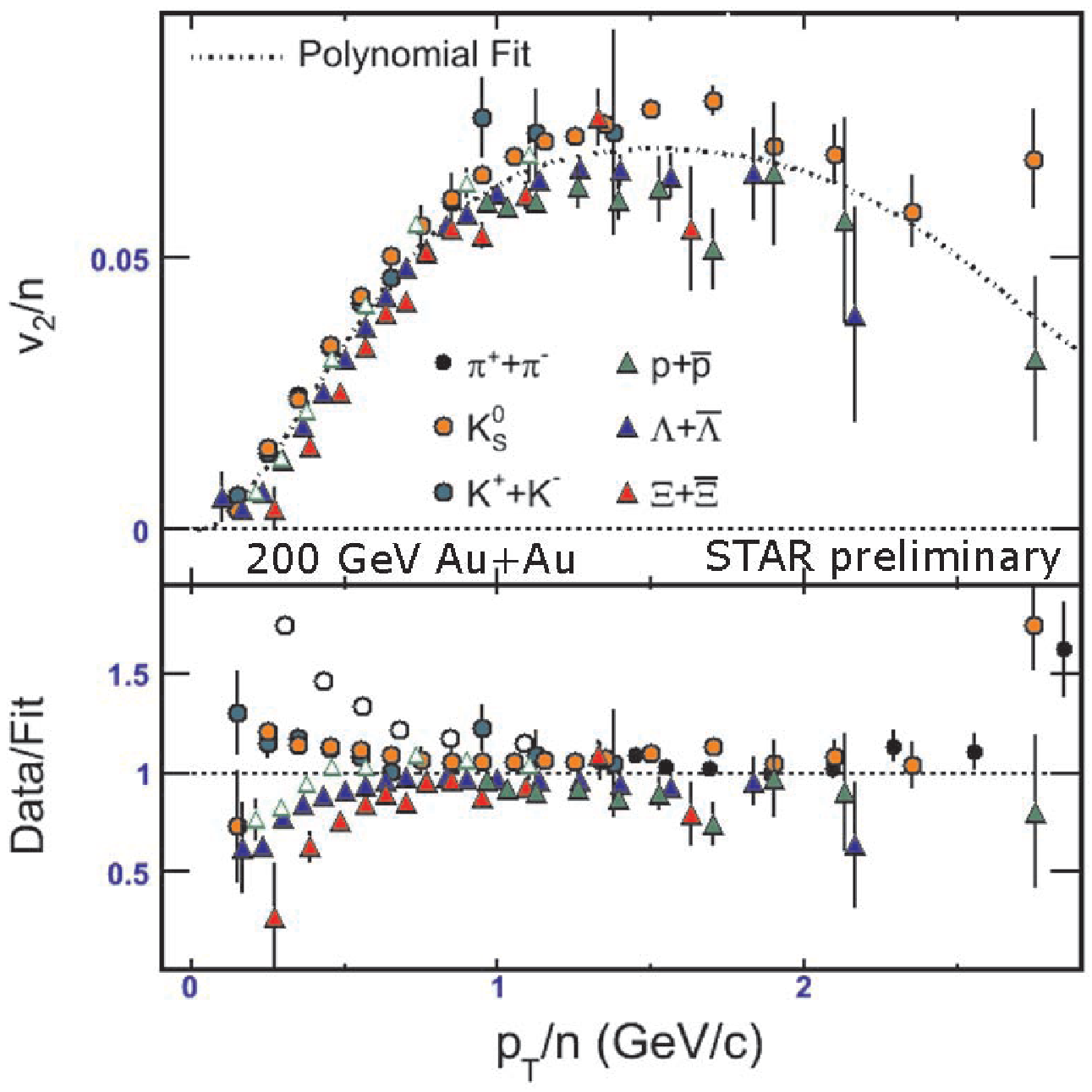}
\vskip -0.8cm
\caption{ $v_2/n$ vs. $p_T/n$; $n$ is the number of valence 
quarks. The bottom panel show the deviations from quark 
number scaling \cite{qm05_flow}.}
\label{fig:quark_num_scaling} 
\end{minipage}
\vskip -0.6cm
\end{figure}

	The universal scaling discussed above, clearly speaks to the validity 
of perfect fluid hydrodynamics over a broad range of the observed data. 
However, such scaling does not provide explicit evidence for the degrees 
of freedom in the flowing matter. For intermediate values of $p_T$, hydrodynamic 
scaling breaks down (cf. Fig.~\ref{fig:v2_pi-k-p_scaled}) and the degrees 
of freedom can reveal themselves. 
Fig.~\ref{fig:quark_num_scaling} shows the results from a recent test for 
quark number scaling of $v_2$ by the STAR collaboration \cite{qm05_flow}.
The lower panel of the figure illustrates the deviations from the 
quark coalescence ansatz used to scale the data shown in the top panel.
The deviations are largest at low $p_T$ where hydrodynamic scaling was 
shown to work best. For intermediate $p_T$s (where the fluid
dynamical picture breaks down) the deviations are quite small, 
suggesting that the active and relevant degrees of freedom are those 
of constituent or valence quarks.   

\section{Conical flow revisited}

Objects moving at supersonic speeds create ``conical flow" behind the shock
front they create. Such flow were initially conjectured to occur in 
cold nuclear matter \cite{early_conical_flow}. However, cold nuclear matter
was later found to be too dilute and dissipative to sustain conical flow. 
The recent discovery of jet quenching at RHIC \cite{jet_quenching} suggests that jets 
created in heavy ion collisions deposit a large fraction, if not all, of their 
energy into the produced matter. If this matter is indeed a strongly coupled Quark-Gluon 
Plasma (sQGP) having a small viscosity, then the current conjecture is that 
this energy should propagate outward in the form of "conical flow" as illustrated 
in Fig.~\ref{fig:mach_cone} \cite{mach_cones}.
Extensive searches for such flow, via two- and three-particle correlation 
functions, are currently underway \cite{3pc_functions,3pc_ajit}.

\begin{figure}[!htb]
\begin{minipage}[t]{0.31\linewidth}
\includegraphics[width=1.\linewidth]{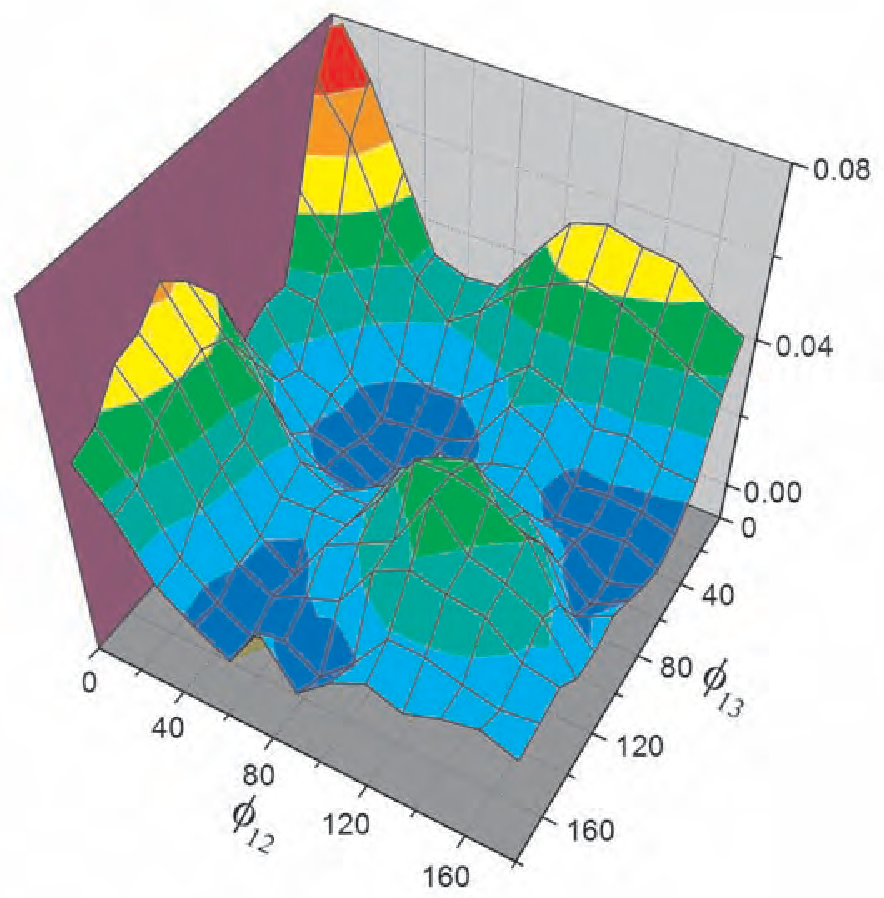}
\vskip -1.0cm
\caption{\small{ Hadron-hadron-hadron correlation 
function \cite{3pc_functions}.}}
\label{3pc_data_hhh}
\end{minipage}
\hskip 0.2cm
\begin{minipage}[t]{0.31\linewidth}
\includegraphics[width=1.\linewidth]{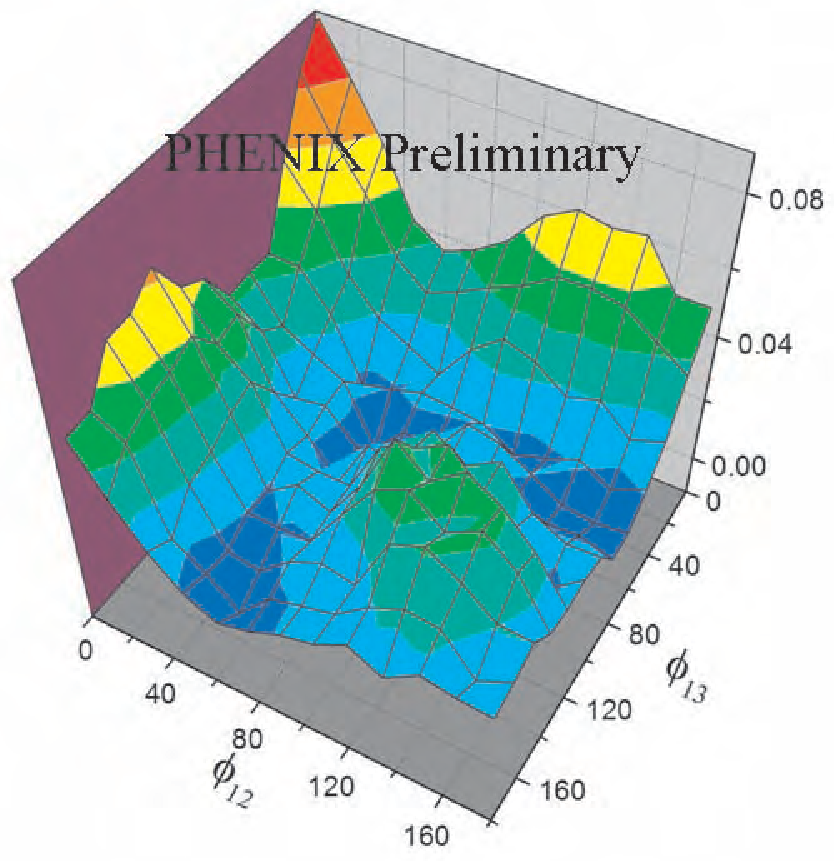}
\vskip -1.0cm
\caption{\small{ Hadron-meson-meson correlation 
function \cite{3pc_functions}.}}
\label{3pc_data_hmm}
\end{minipage}
\hskip 0.2cm
\begin{minipage}[t]{0.31\linewidth}
\includegraphics[width=1.\linewidth]{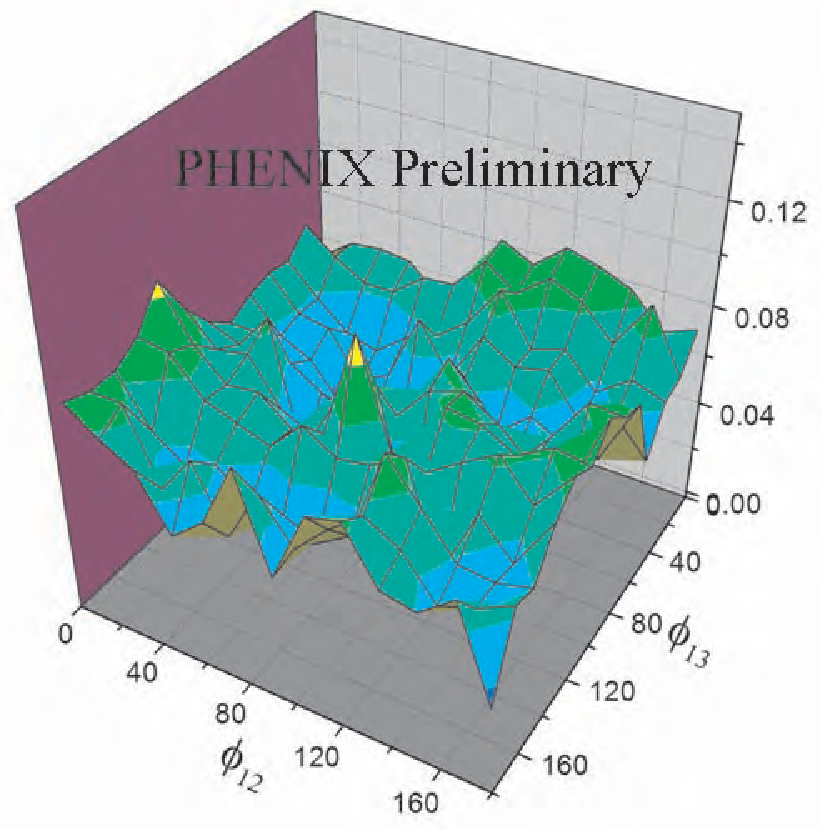}
\vskip -1.0cm
\caption{\small{ Hadron-baryon-baryon correlation 
function \cite{3pc_functions}.}}
\label{3pc_data_hbb}
\end{minipage}
\hskip 0.2cm
\end{figure}
Figs.~\ref{3pc_data_hhh}~-~\ref{3pc_data_hbb} show three-particle $\Delta\phi$ jet correlation 
surfaces ($\Delta\phi_{1,2}$ vs. $\Delta\phi_{1,3}$) for a trigger hadron from the 
range $2.5< p_T <$4.0 GeV/c (hadron \#1) and two associated hadrons from 
the range $1.0< p_T <$2.5 GeV/c (hadron \#2 and \#3) \cite{3pc_ajit}. Results are shown for 
the centrality selection 10-20 {\%}. They show a strong dependence on the flavor (PID) 
of the associated particle and clearly do not follow the expected patterns for 
a ``normal jet" \cite{3pc_ajit}. Instead, they provide rather compelling evidence for strong 
modification of the away-side jet. Further detailed quantitative investigations are 
required to firm up whether or not the tantalizing qualitative features evidenced in 
Figs.~\ref{3pc_data_hhh} and \ref{3pc_data_hmm} confirm the presence of conical flow.

\section{Epilogue} 

In summary, the gross elliptic flow patterns observed at RHIC provide compelling 
evidence for the production and rapid thermalization 
(less than 1 fm/c after impact) of nuclear matter having an 
energy density well in excess of an order of magnitude above the 
critical value required for deconfinement. 
The universal and fine structure scaling properties of this flow, indicate 
that this new form of matter behaves like a perfect liquid. The quark 
number scaling observed at intermediate $p_T$s (when the fluid
dynamical picture breaks down), indicates that the relevant degrees of freedom
are those of constituent or valence quarks.
These properties are consistent with those of a strongly coupled plasma 
with essentially perfect fluid-like properties. 
Systematic experimental and theoretical efforts are currently underway to 
quantitatively pin down the thermalization time, EOS and transport properties 
of this plasma.
\end{document}